\documentclass[12pt,harvard]{iopart}
\usepackage{iopams}  

\usepackage[final]{graphicx}

\usepackage{verbatim}
\usepackage{algorithm}
\usepackage{algpseudocode}
\usepackage{harvard}

\begin{document}

\title[Running title: The
 Time Series Tendency]{
Defining a Trend for a Time Series Which Makes Use of the Intrinsic Time-Scale Decomposition}

\author{Juan M. Restrepo$^{1,2,3}$,  Shankar Venkataramani$^{1,2}$, Darin Comeau$^{2}$ \& Hermann Flaschka$^{1,2}$}

\address{$^1$Mathematics Department, $^2$ Program of Applied Mathematics, $^3$Physics Department, University of Arizona, Tucson AZ 85716 USA}
\ead{restrepo@math.arizona.edu}
\begin{abstract}
We propose criteria that define
 a  trend for time series with  inherent multi-scale features. 
 We call this trend  the {\it tendency} of a time series.
 The  tendency is defined empirically by a set of criteria and 
 captures the large-scale temporal variability of the original signal as well as the most frequent events in its histogram.  Among other 
 properties, the tendency has a  variance no larger than that of the original signal; the histogram of the difference between the original signal and the tendency is as symmetric as possible; and  with reduced complexity, the tendency captures
essential   features of the signal.
  
To find the tendency we first use the
Intrinsic Time-Scale Decomposition  (ITD)
of the signal, introduced in 2007 by Frei and Osorio, to produce a set of candidate
tendencies. We then apply the criteria to each of the candidates 
to single out the one that best agrees with them.

While the criteria for the tendency are independent of the signal
decomposition scheme, it is found that the ITD is a simple
and stable methodology, well suited for multi-scale signals.
The ITD is a relatively new decomposition and little is known
about its outcomes. In this study we 
 take the first steps towards a probabilistic
model of the ITD analysis of random time series. This analysis yields
details concerning the universality and scaling properties of
the components of the decomposition. 

 \end{abstract}

\vspace{2pc}

\submitto{\NJP}

 \section{Introduction}

Finding the trend of a time series is a fundamental analytical task. 
To varying degrees, the definition of the term ``trend" is
dependent on the methodology used to compute it. Some trending
strategies are optimal and thus very attractive because the optimality criteria  provide
mathematical constraints with which to interpret the time series trend.   Not all optimal trends deliver useful trends. An example of a non-optimal trend is the Hodrick-Prescott Filter (see \cite{hpfilter}), which is widely used  in econometrics. 
This paper proposes a new empirically-defined trend for an inherently multi-scale, finite time series. 

In an econometric context, the trend is often used to capture the longer time scale structure of the markets, by filtering out high frequency events that might be more relevant to shorter time scale changes. It is used this way in the physical sciences as well.
Our interest in this topic was motivated by the problem of trend determination in a geoscience context, where, in addition to the removal of biases, one is often confronted with the necessity of analysing signals with the aim of recovering structural aspects of the signal that can be captured or explained by physical models.

Geoscience problems often involve multi-physics and other sources of complexity which manifest themselves in a time series with a rich variety of time scales. There is no rigorous definition of ``multi-scale" signals, but in the physical modeling community the adjective is applied to signals that are the result of the coupling of inherent degrees of freedom (sometimes given by spectral components). Furthermore, it is frequently the case in geoscience that the time series to be analysed is short, of length much shorter than the number of degrees of freedom of the system that generated the series; sometimes  too short to be amenable to law-of-large numbers statistics.

 The procedure we propose to find the tendency is a two-stage process: we first decompose the signal in a series of time series of progressively lower complexity, 
 and then we
  apply a set of criteria to these and single out the decomposition mode that best 
  satisfies the criteria.
 This mode is declared to be the tendency of the signal.
 
While we employ the
Intrinsic Time Decomposition (ITD) of \cite{itd}, the
strategy could be applied to other algorithms,
such as the Emprical Mode Decomposition (EMD),
\cite{huangemdfirst,huangemd,wuhuanglongpeng}. In fact, a similar procedure has been proposed in connection with EMD (see \cite{mbf11,mbf13}): the EMD modes are
calculated, and a time series representing the trend is built according to
certain prescriptions.

Most everything that is known   to date about the  ITD will be reviewed in Section \ref{itd}; the algorithm  that performs the decomposition appears in the Appendix. In Section \ref{itd}
we summarize results of
 computer experiments on random signals that suggest
 certain probabilistic and scaling features in the  ITD decomposition. 
 In Section \ref{diffusion}
we initiate  a mathematical analysis of these numerical results.
This analysis  might be applicable to the EMD
 its variants,  such as those proposed by  \cite{houemd} and \cite{tmp}. We also use computer experiments to draw attention to 
 the influence of boundary conditions on the outcomes, focusing only on the ITD.  Boundary effects are seldom highlighted in the EMD and ITD papers, but we find
that for some signals, the
boundary conditions can have a significant effect on the outcomes and thus on the 
construction of the tendency from the
components of the decomposition. This discussion appears in 
 Section~\ref{endconditions}.

 Section \ref{tendency} introduces the criteria that are used to pick the ITD mode
 that is declared to be the tendency. 
 The criteria are empirically-based
 notions of signal information whose implementation is discussed in  
Section \ref{examples}. In that section,
 we illustrate the application of  the two-step process for
 finding the tendency on deterministic and random signals as well as on
 real geophysical signals.   In the latter group, we will feature an analysis of the 
post-industrial temperature  record in Moscow (available from \cite{giss}). 
Rahmstorf and Coumou (see  \cite{rahmstorfextremetemps}) set out to determine whether the extreme Moscow summer temperatures of 2010 were  outlier samples of climate or the result of an ever warming Earth.  Their analysis of extreme events depends on the proper determination of a sensible long-time trend, or ''climate.'' In Section \ref{disc} we summarize
the outcomes of the analysis and the outcomes of the tendency calculations, and take
the opportunity to compare,
in general terms, the ITD tendency and the EMD trend.

\section{The Intrinsic Time-Scale Decomposition}
\label{itd}
 The Intrinsic Time-Scale Decomposition (ITD) is a purely algorithmic, non-lossy 
iterative decomposition of 
a time series $\{Y(i)\}_{i=1}^N$.  At the first stage, the signal is decomposed into a
{\it proper rotation} $R^1(i)$, an oscillating mode
in which maxima and minima are positive and negative,
respectively, and a residual $B^1(i)$
called {\it baseline} .

The baseline
$B^1$ is now decomposed in the same fashion, producing a proper rotation $R^2$
and a baseline $B^2$, and so on. The process stops when the resulting baseline has only two extrema, or is a constant.

If there are $D$ steps altogether, the decomposition has the form
\begin{equation}
B^0(i):=Y(i) = B^D(i) + \sum_{j=1}^{D}R^j(i),\ i=1,...,N.
\label{itddecomp}
\end{equation}
Rotations and baselines satisfy the relation
\begin{equation}
B^j(i) = B^{j+1}(i) + R^{j+1}(i),\ i=1,...,N; j=0,...,D.
\label{relation}
\end{equation}
Parenthetically, we note that in the algorithm as described in
\cite{itd} there is one, and only one,
adjustable parameter denoted by $\alpha$, which has been set to  $\alpha=1/2$ in ouer study. 

In general, the rotation signal at the $j^{\mathrm{th}}$ level will be "noisier" than 
the rotation signal at $(j+1)^{\mathrm{th}}$.  The proper rotations are not orthogonal; moreover, the decomposition is not linear, in the sense that a decomposition of the sum of
time series is not equal to the sum of the decompositions of each of the signals.

Let  $\{\tau^j_k\}$, $k=1,2,..,K$ be the times at which
the extrema of $B^j(i)$ occur.
(In the event that there 
are several successive data points with the same extremal value, we take $\tau^j_k$ to 
correspond to the time of the rightmost of these extremal values). The  baseline $B^{j+1}(i)$ is  constructed by
a piecewise linear formula: in the interval $i \in (\tau^j_k, \tau^j_{k+1}]$, between successive extrema,
\begin{equation}
B^{j+1}(i)  = B^{j+1}_k + \frac{(B^{j+1}_{k+1}-B^{j+1}_k)}{(B^j_{k+1}-B^j_k)}(B^j(i)-B^j_k), 
\label{bi}
\end{equation}
where the {\it knots} are $B^j_k :=B^j(\tau_k)$. The formula that generates the knots is
\begin{equation}
B^j_{k+1}:=B^j(\tau_{k+1}) = \frac{1}{2} \left[ B^j_{k-1} +   \frac{(\tau^j_{k}-\tau^j_{k-1})}{(\tau^j_{k+1}-\tau^j_{k-1})}(B^j_{k+1}-B^j_{k-1}) \right] + \frac{1}{2}  B^j_{k}.
\label{bk}
\end{equation}
The construction guarantees that  the residual function
\begin{equation}
R^{j+1}(i) = B^{j}(i) - B^{j+1}(i), \quad i=1,2,...,N,
\label{ri}
\end{equation}
is monotonic between adjacent extrema.
Figures illustrating the construction may be found in
\cite{itd}. 

One must also decide on a
boundary condition
at the two ends. The effects of different choices will be discussed in
Section~\ref{endconditions}.  
We shall interpret the end points as extrema, and take the corresponding baseline
knots to be averages of the first and last pair of extrema, 
$B^j_1= B^j(1)$,  and $B^j_{K^j}= B^j(N)$:
\label{bcs} \\
\begin{equation}
B^{j+1}_1 = \frac{1}{2} (B^j_2+B^j_1)  \quad \mbox{ \ \   and } \quad
 B^{j+1}_{K^j} = \frac{1}{2} (B^j_{K^{j-1}}+B_{K^j}). \label{freebc}
\end{equation}
These will be  called  {\it free} boundary conditions.

The decomposition ends when $j=D$, which is  when a proper rotation cannot be constructed from this last baseline. Baseline $B^D(i)$ will only have two knots: $B_{k=1,2}^D$, the two end points.

We now state a few important properties of the ITD decomposition, largely following \cite{itd}.
\begin{enumerate}
\item 
The baselines given in  (\ref{bi}) can be rewritten as a convex combination; {\it viz.}, 
$
B^{j+1}(i) = (1-s^j_k(i) ) B^j_k + s^j_k(i) B^j_{k+1}, \qquad s^j_k(i) = \frac{B^j(i)-B^j_k}{B^j_{k+1}-B^j_k},
$
where $s^j_k(i) \in [0,1]$, and $j=0,1,..,D$; 
\item The knots, (\ref{bi}),   at level $j+1$ can also be written as 
\begin{equation}
B^{j+1}_k = \frac{1}{2} B^j_k +
\overline{B_k^j} + \frac{1}{2}\frac{\tau^j_{k+1}+\tau^j_{k-1} - 2 \tau^j_j}{\tau^j_{k+1}+\tau^j_{k-1}}\left( B^j_{k+1}-B^j_{k-1} \right),
\label{alt3}
\end{equation}
 where the overline indicates average of nearest neighbors;
\item The ITD decomposition is ambiguous with regard to handling the end points of a finite time series, and thus,
different end conditions can generate different ITD decompositions. See Section~\ref{endconditions}; 
\item 
The baseline extraction step can be thought of as a nonlinear  
operator ${\cal L}$,
homogeneous with respect to independent rescaling of the abscissa and also the ordinate:
 $B^{j+1}= {\cal L} B^j$ and $R^{j+1} = (1-{\cal L}) B^j$. 
 \item The $B^j$ and $R^j$ are monotonic between successive extrema, since 
they are obtained, in succession, through linear transformations; 
\item It follows from the above property, that the $\ell_2$-norm of $B^j$ is similar (to within a
 constant) to the $\ell_2$-norm of
an approximation of the same signal, built by connecting extrema with piece-wise linear segments; 
\item  The extrema of $B^j$ are inflection points or extrema of $B^{j+1}$; 
\item  Between extrema of $B^j$, $B^{j+1}$ has the same smoothness as $B^j$;
\item At extrema of $B^j$, $B^{j+1}$ will be continuous and differentiable, but not always twice differentiable;
\item  $R^{j+1}(i)$ will have extrema at the same locations as $B^j(i)$. 
\end{enumerate}

\subsection{Random signals} \label{sec:random}
We want to understand some basic features of the ITD before we try to
extract the tendency of a realistic signal.
Since we use the ITD to strip random noise from a time series, it is appropriate to 
begin by applying the method to purely random signals, and furthermore,
since the ITD extracts
the rotation components in order of increasing wavelength, we start with a
random series in which every point is a local extremum.  As 
already mentioned above,
we study the scaling properties of the wavelengths of the baseline $B^j$,
numerically in this section, and analytically in the next.

Our time series has the form
\begin{equation}
Z(i) = (-1)^i | z_i|, \quad i = 1,2,...,N.
\label{zeq}
\end{equation}
The random variables $z_i$ are drawn from a normal  ${\cal N}(0,\sigma^2)$.
Definition (\ref{bk}) for the baseline at the initial step becomes
$$ 
B^1_k = \frac{1}{4} (Z_{k-1} + 2 Z_k + Z_{k+1}), \quad k=1,2, \ldots, N.
\label{bkz}
$$
The corresponding proper rotation is
$$
R^1_k = Z_k - B^1_k = -\frac{1}{2} (Z_{k-1} - 2 Z_k + Z_{k+1}).
\label{rkz}
$$ 
By periodizing and taking $N$ even data points, the ratio of the discrete  Fourier transform of this $B^1$ to $Z$ yields
\begin{equation} \label{bkzFT}
\frac{\hat B^1}{\hat Z} = \frac{1}{2} (1+\cos \omega),
\end{equation}
where $\omega = 2 \pi \nu/N$, and $0 \le \nu \le N/2$, the integer frequency.
Similarly, the ratio of the transform of $R^1$ and $Z$ gives
\begin{equation} \label{rkzFT}
\frac{\hat R^1}{\hat Z} = 1-\frac{\hat B}{\hat Z}=\frac{1}{2}(1-\cos \omega).
\end{equation}
One sees that $B^1$ and $R^1$ are obtained by
convolving the signal~$Z$ with a low-pass, resp. high-pass, filter.
If $Z$ were a discrete sinusoid with a highest
frequency of $N/2$,  $R$ would be an exact copy of $Z$, while
$B$ would be zero, and there would be no further decompositions.
Generally, however,
 the averaging operator (\ref{bkz})
 will tend to smooth features that appeared in the original signal, and
thus, the resulting baseline will generally have a different distribution of extrema than the original
signal, see Section~3.
\begin{figure}[hbt]
\centering
(a)\includegraphics[width=5.5cm,height=0.75in]{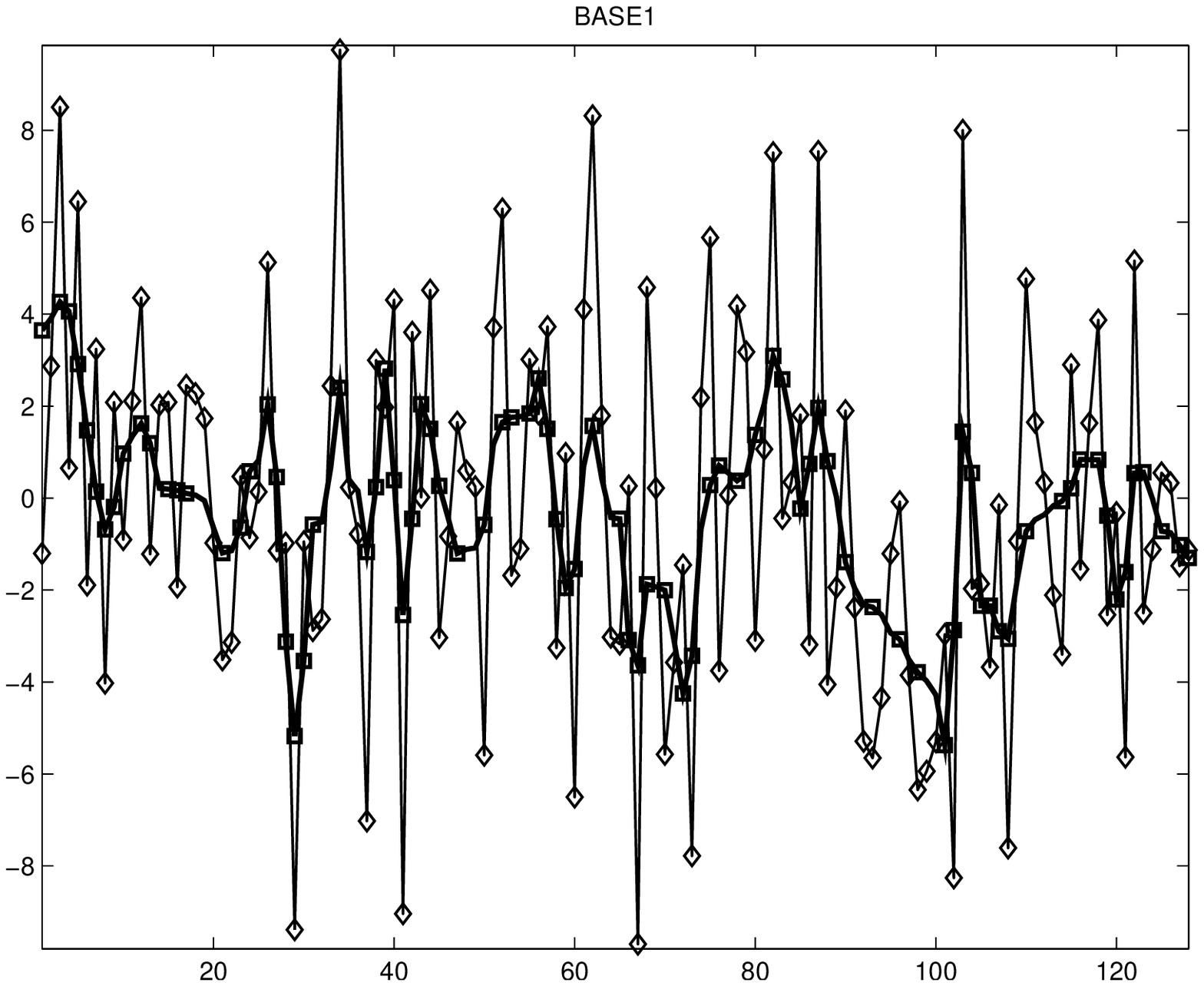}
(b)\includegraphics[width=5.5cm,height=0.75in]{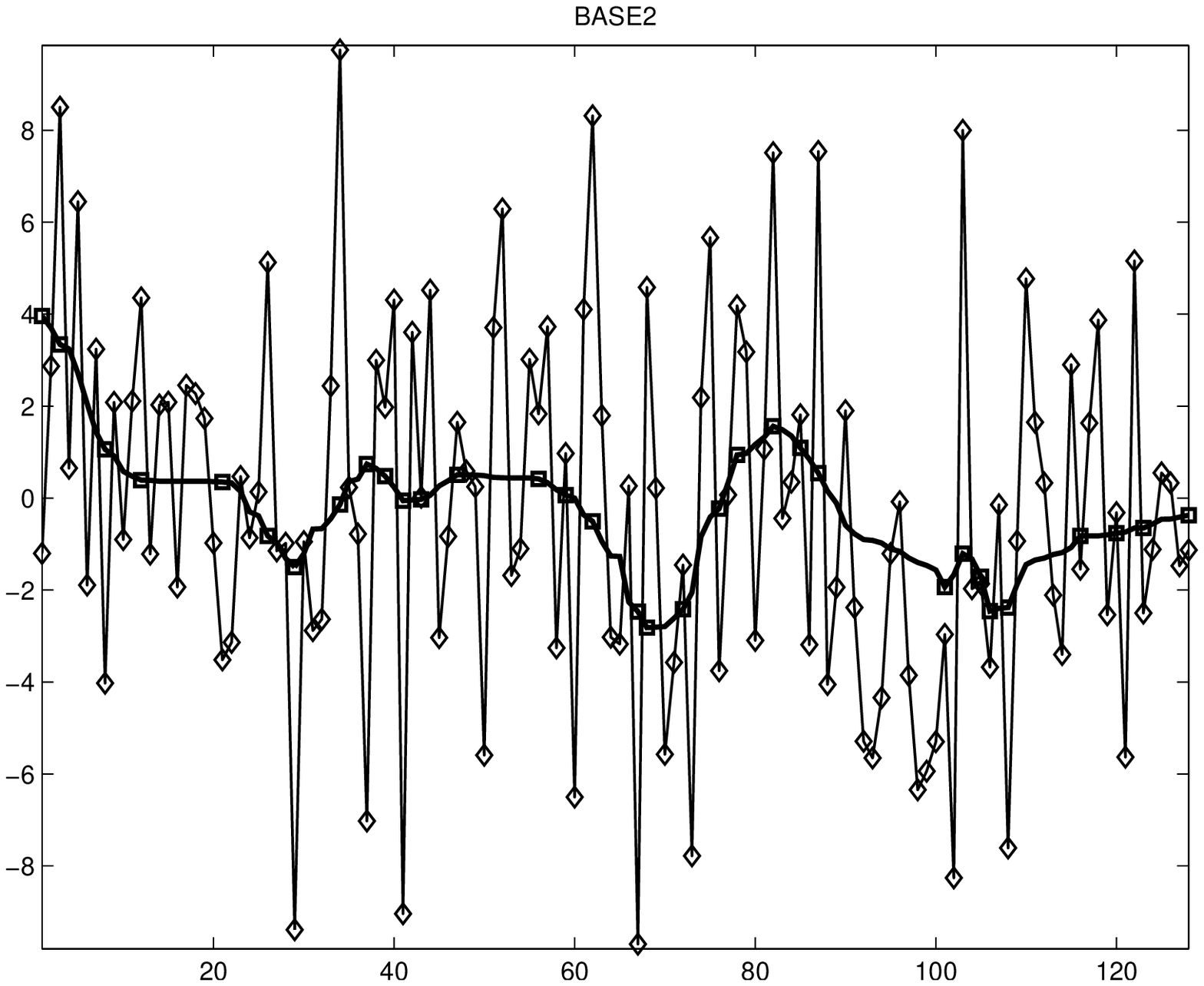}
(a')\includegraphics[width=5.5cm,height=0.75in]{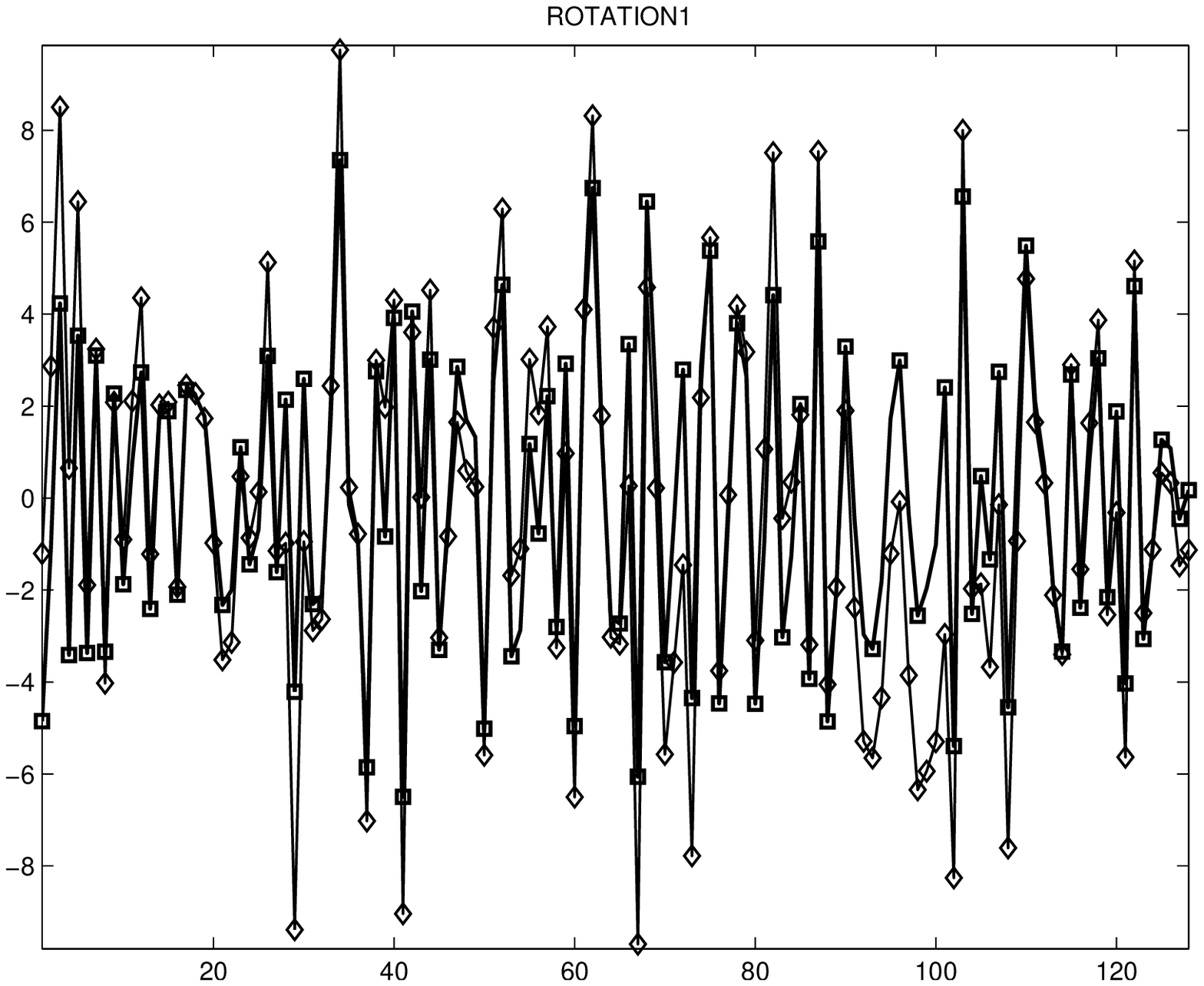}
(b')\includegraphics[width=5.5cm,height=0.75in]{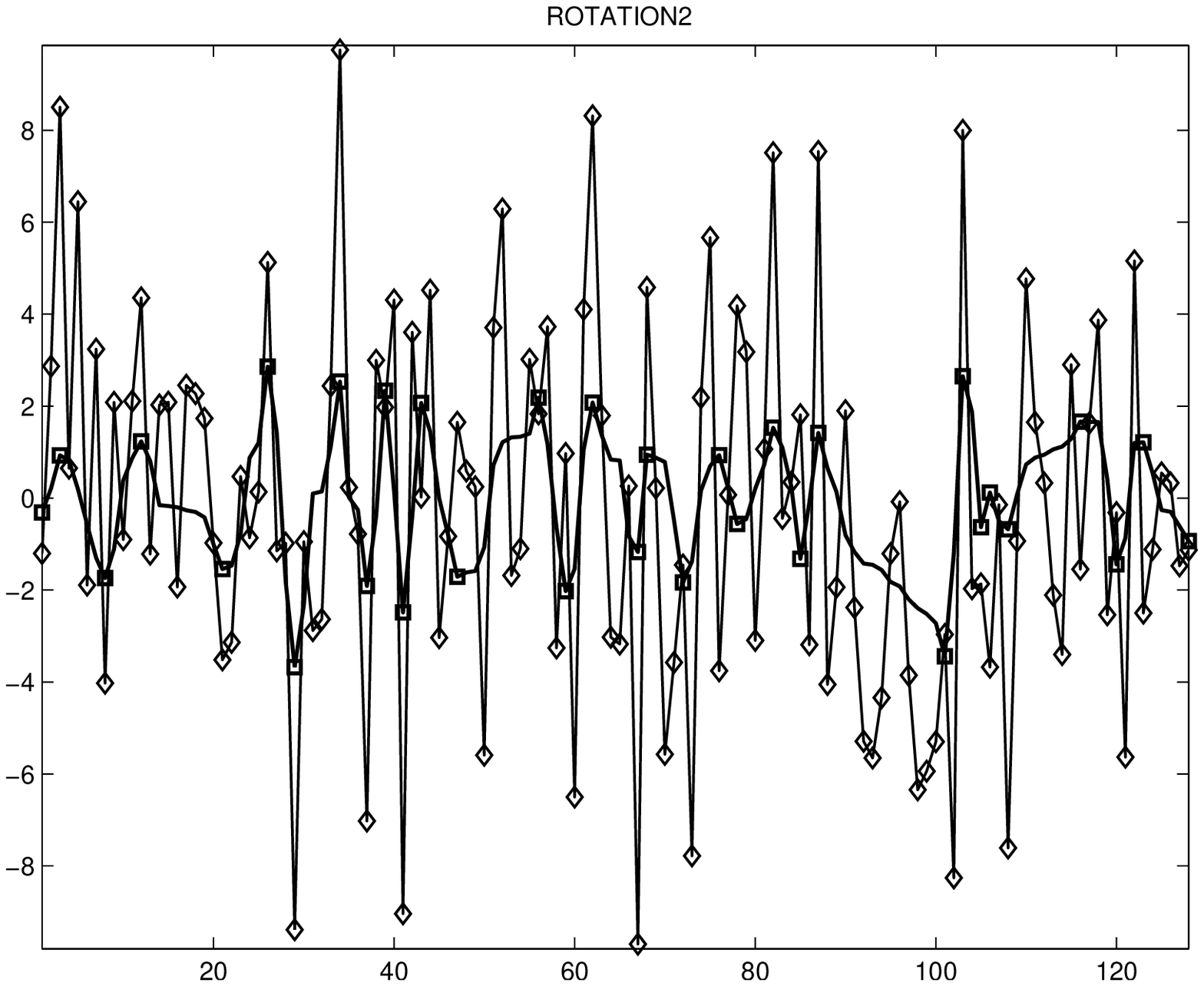}
(c)\includegraphics[width=5.5cm,height=0.75in]{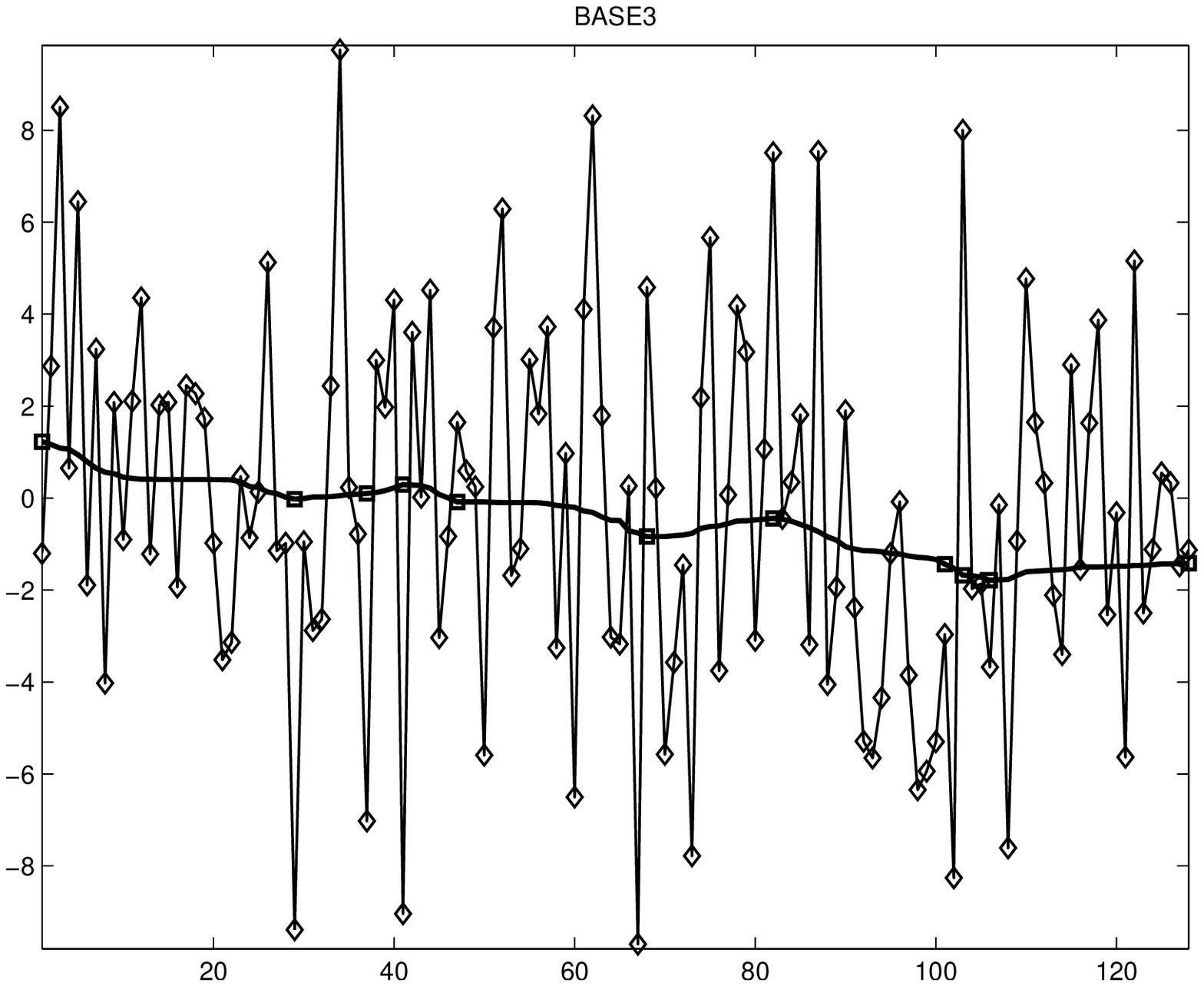}
(d)\includegraphics[width=5.5cm,height=0.75in]{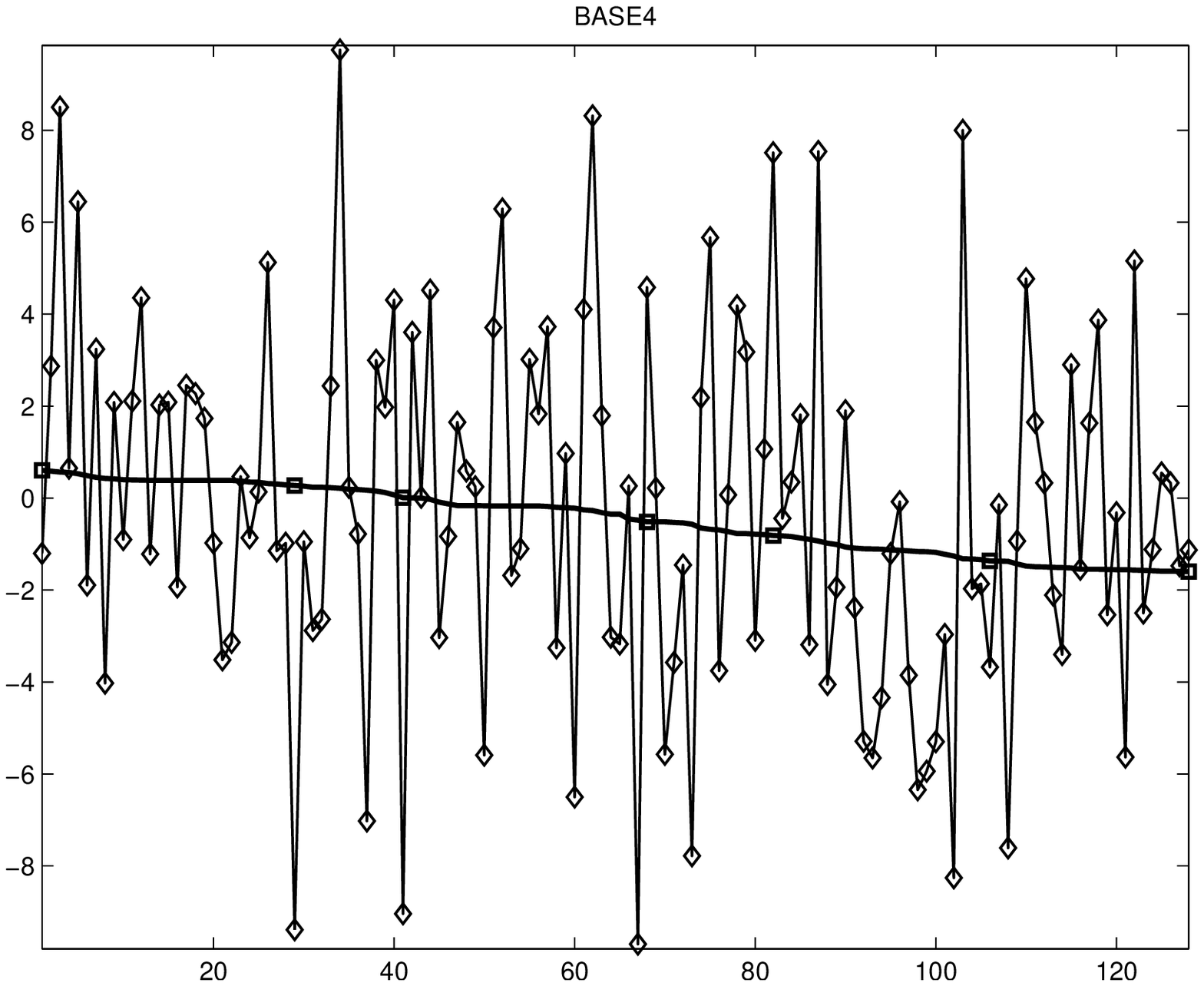}
(c')\includegraphics[width=5.5cm,height=0.75in]{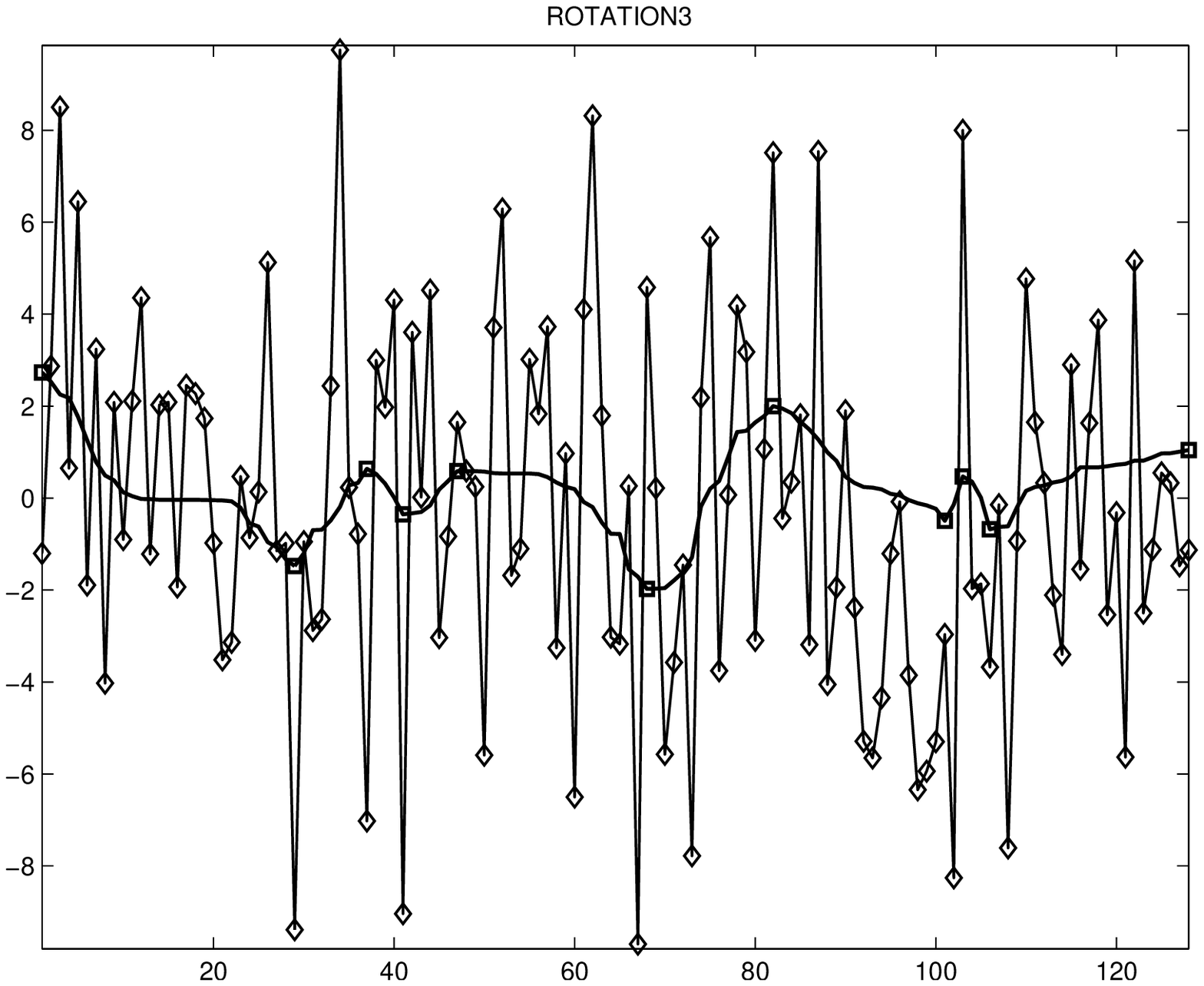}
(d')\includegraphics[width=5.5cm,height=0.75in]{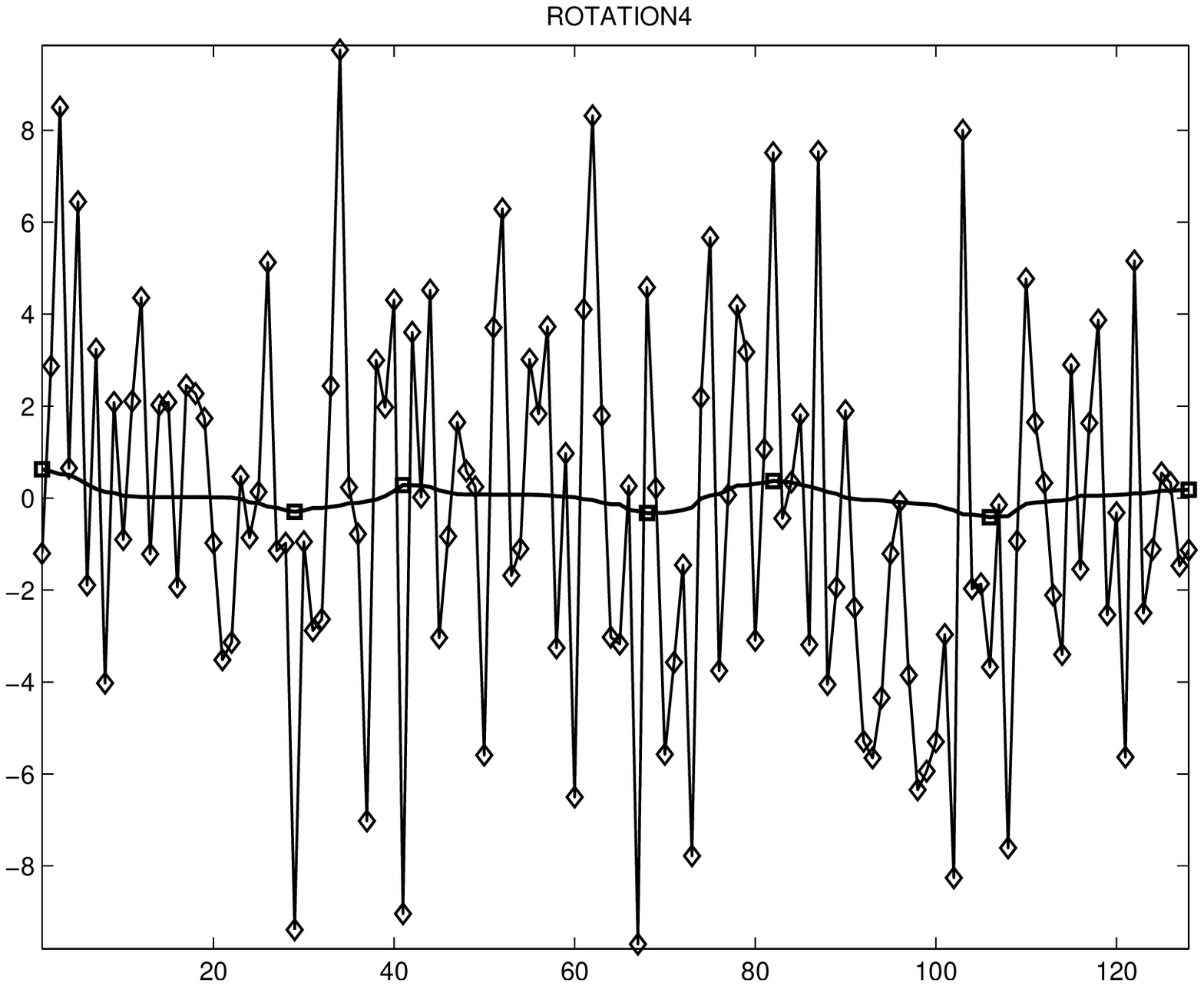}
\caption{
(a) Signal $Z$, as in (\ref{zeq}), 
with $N=128$ and $z_i$  from ${\cal N}(0,4)$; (b), (c), (d) are the first three baselines, and (a')-(d') the corresponding
rotations.
}
\label{fg.BRscaling}
\end{figure}
Parenthetically we note that the Hodrik-Prescott filter (see \cite{hpfilter}), used in econometrics to find the large-scale trend of financial data, produces a trend $H$ with 
a transfer function
\[
\frac{\hat H}{\hat Z} =  \frac{4 \lambda (1-\cos \omega)^2}{1+4\lambda  (1-\cos \omega)^2},
\]
where $\hat H$ is the Fourier transform of the filter output, and $\lambda$ is a free parameter.  This filter is a windowed low pass filter, capable of handling data from a non-stationary process, however, it is hard to make sense of its outcome if the time series is not at least $I(2)$ (non-stationary and must be differenced twice to obtain stationarity).

\begin{figure}[hbt]
\centering
(a)\includegraphics[height=4.cm,width=3.5cm]{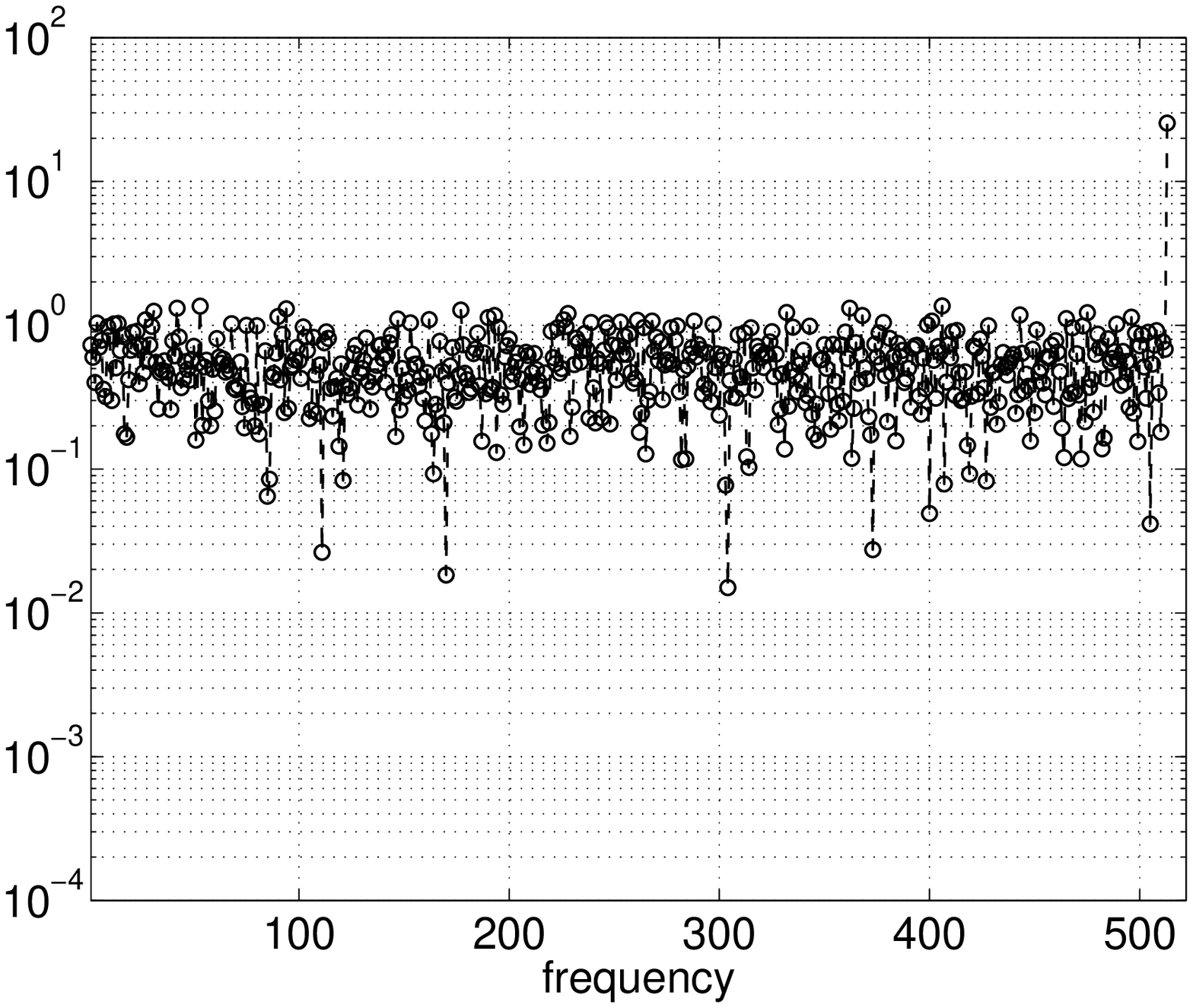}
(b)\includegraphics[height=4.cm,width=3.5cm]{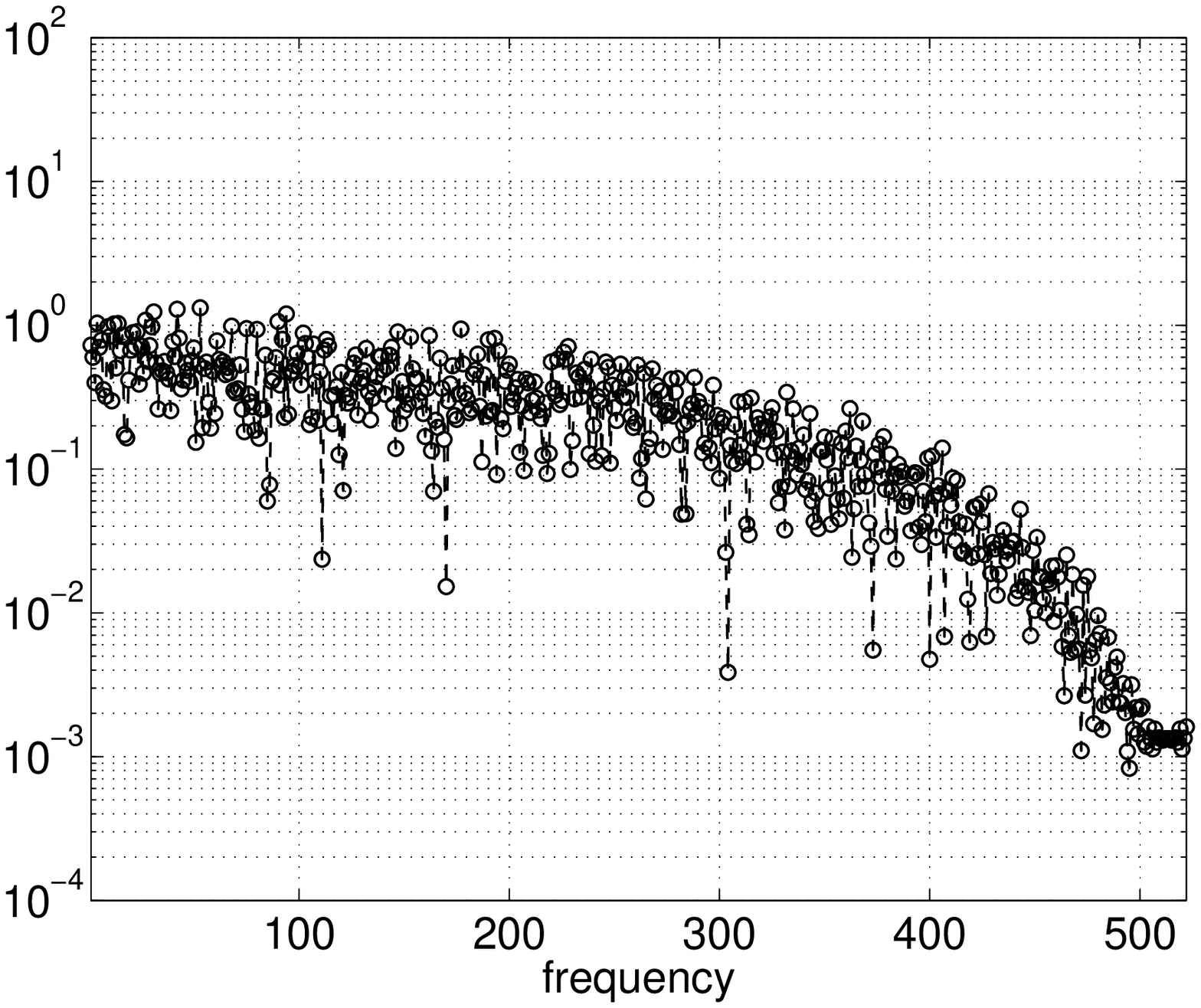}
(c)\includegraphics[height=4.cm,width=3.5cm]{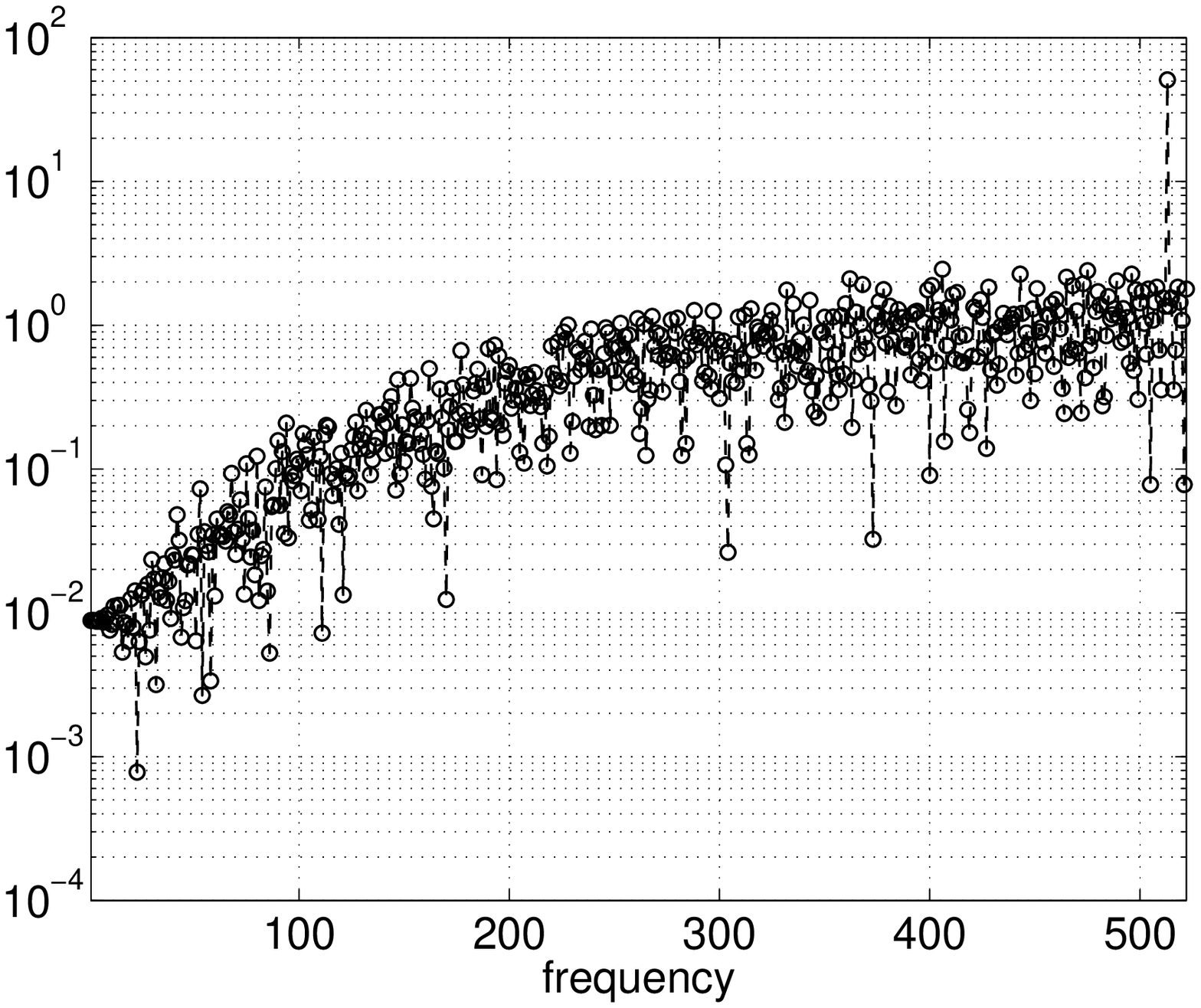}
\caption{(a) Energy spectrum  of  $Z$, as in (\ref{zeq}), normalized to $\|Z\|_2=1$.
$N=1024$, and $z_i$ drawn from 
${\cal N}(0,4)$. Normalized spectra of resulting
$B^1$ and $R^1$ are in panels (b) and (c),
respectively. Note that the original signal has most of its energy concentrated in the highest frequency, $512$.}
\label{fg.spectral} 
\end{figure}

\begin{figure}[hbt]
\centering
(a)\includegraphics[height=3cm,width=5.3cm]{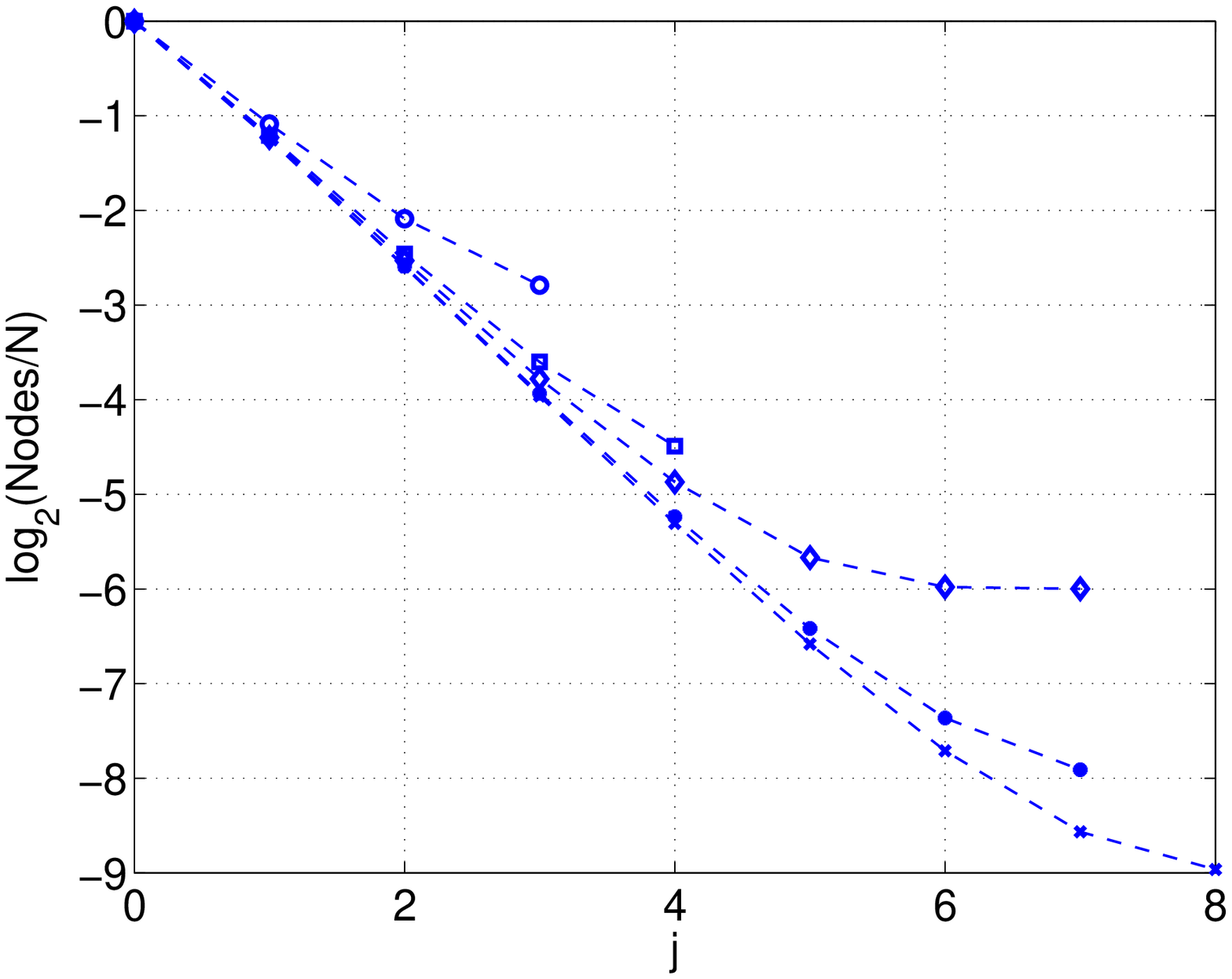}
(b)\includegraphics[height=3cm,width=5.3cm]{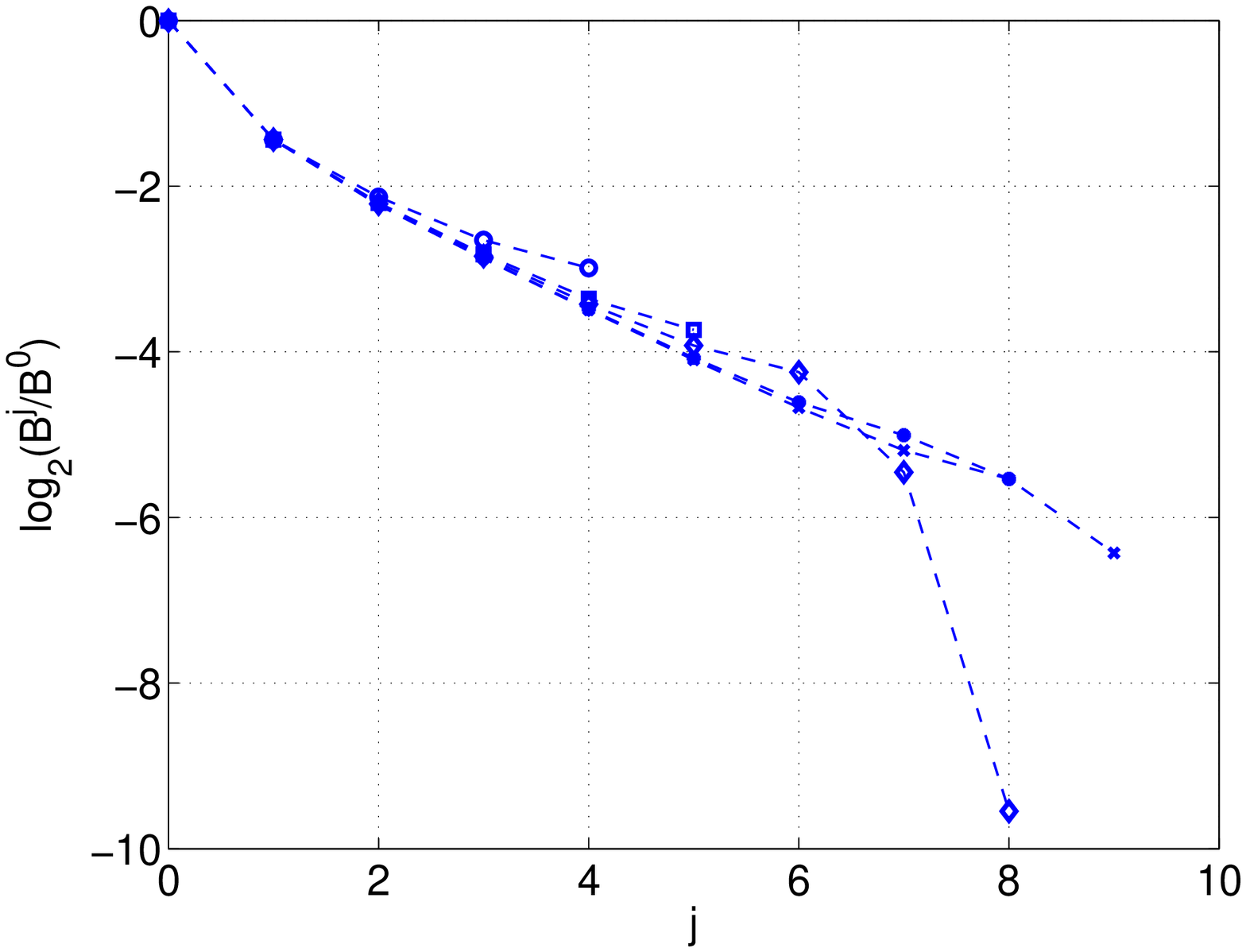}
\caption{Ensemble average of the ITD of $Z$ as per (\ref{zeq}), with $z_i$ drawn from a Normal, with variance $\sigma^2=4$, as function of~$j$.
Mean values at each $j$ of 50,000 realizations of $Z$ and their analyses. The length of the signals was  $N=16,64,128,512,1024$. (a) The log$_2$  of the 
mean number of extrema, of the baselines,
normalized to $N$; 
(b) log$_2(\mbox{mean}\|B^j\|_2/\mbox{mean}\|Z\|_2)$. 
The total number of $j$ levels in  the ITD decomposition of random signals is of order $\log N$.
}
\label{fg.allmaxN}
\end{figure}

\begin{figure}[hbt]
\centering
(a)\includegraphics[width=5.3cm,height=1in]{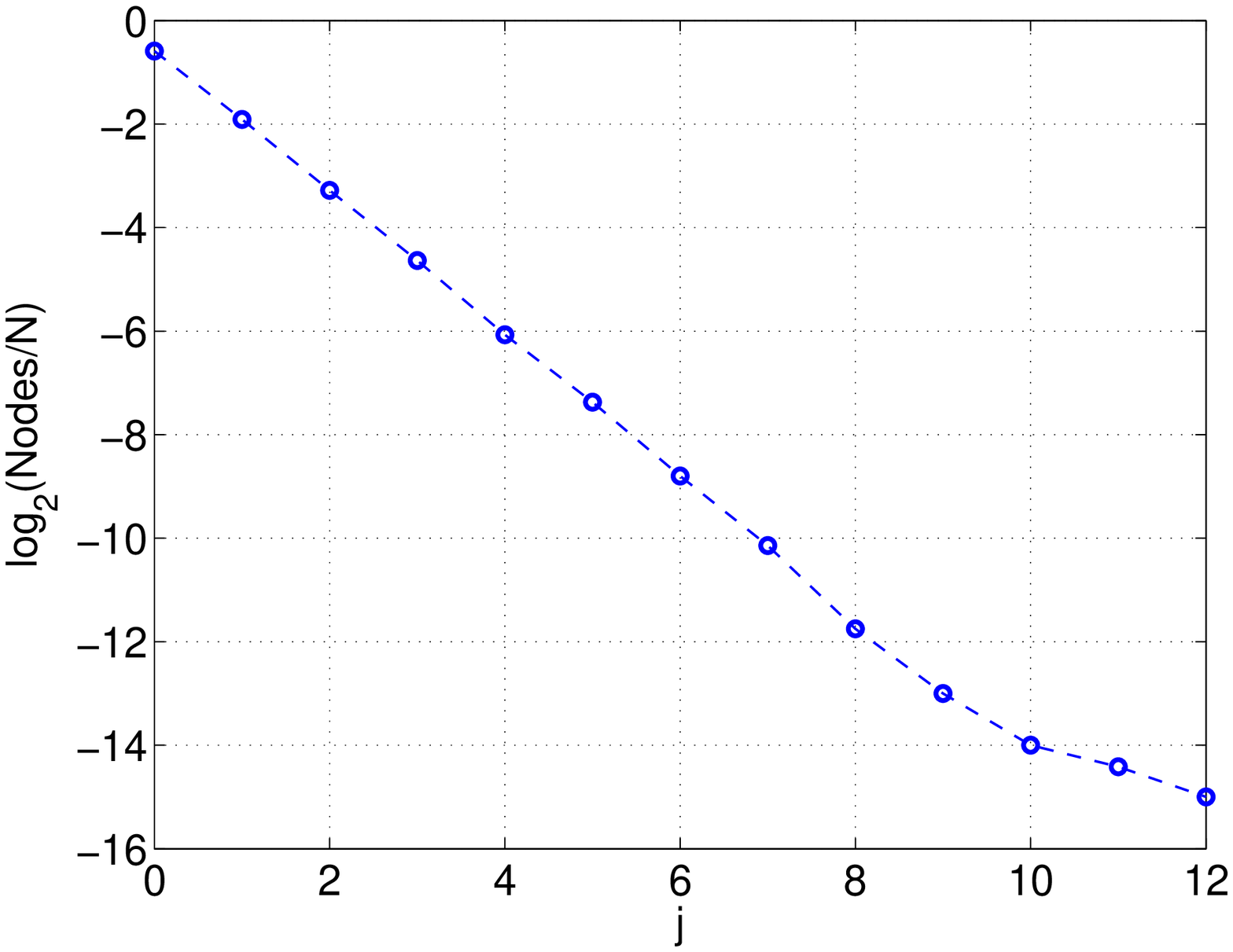}
(b)\includegraphics[width=5.3cm,height=1in]{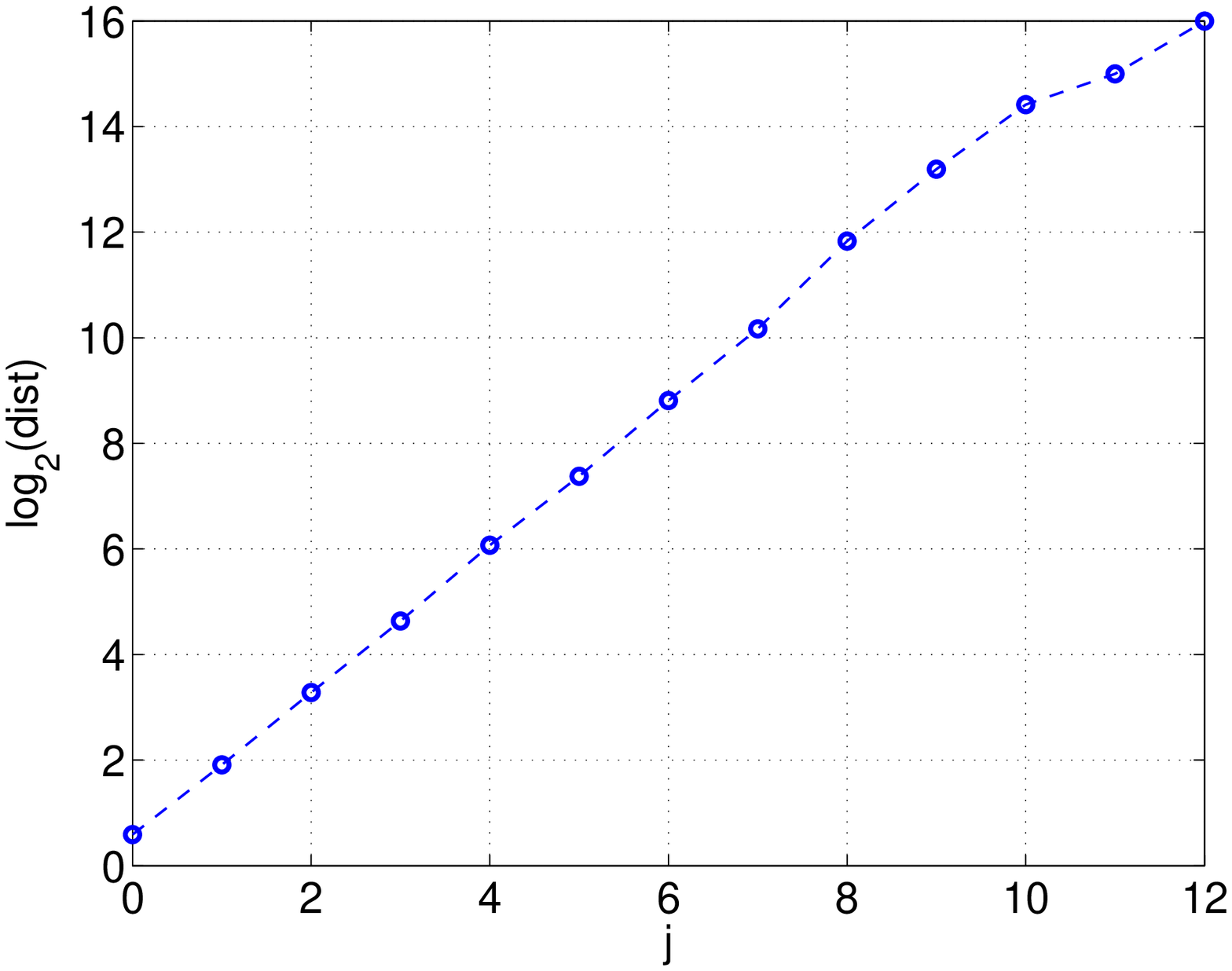}
(c)\includegraphics[width=5.3cm,height=1in]{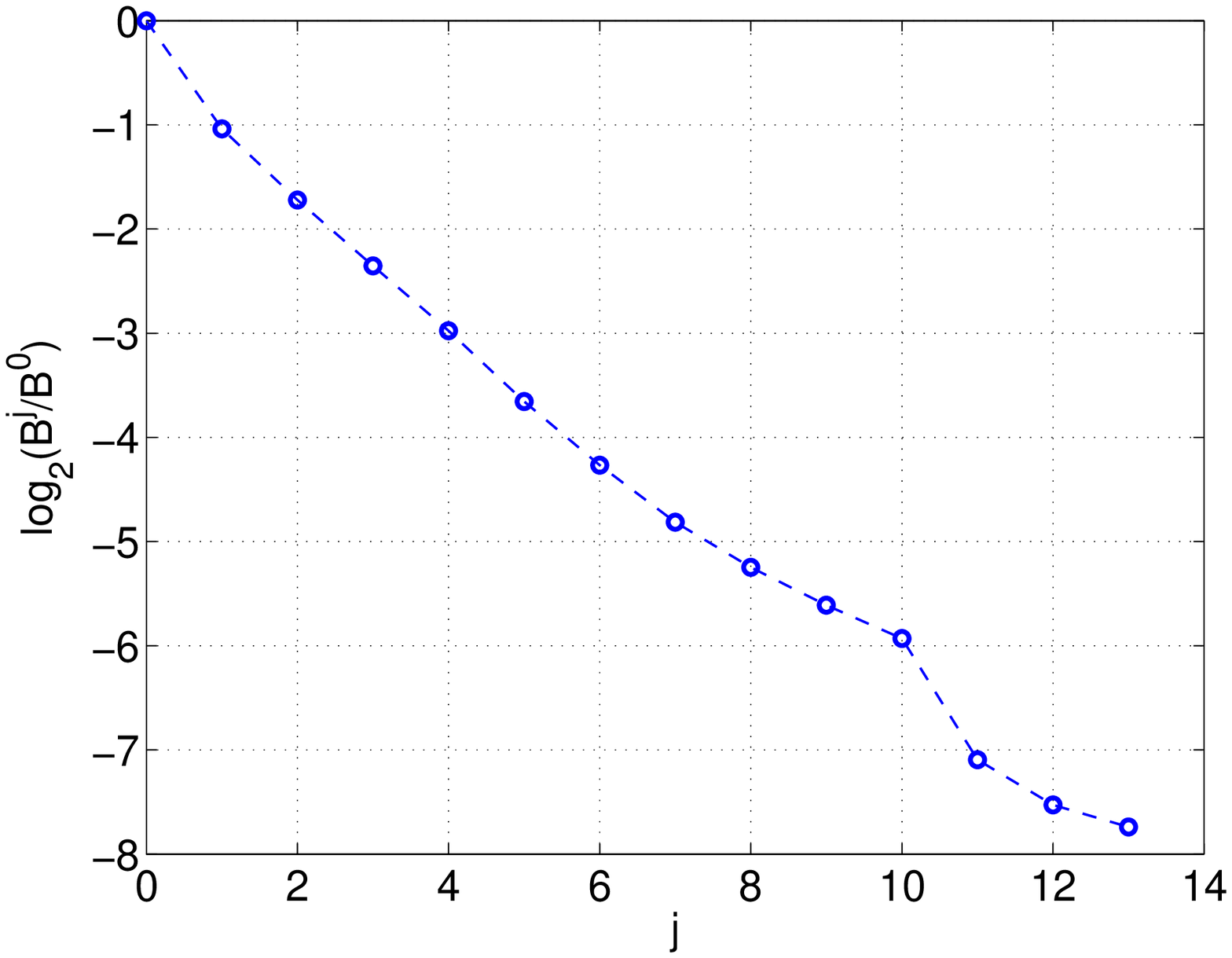}
(d)\includegraphics[width=5.3cm,height=1in]{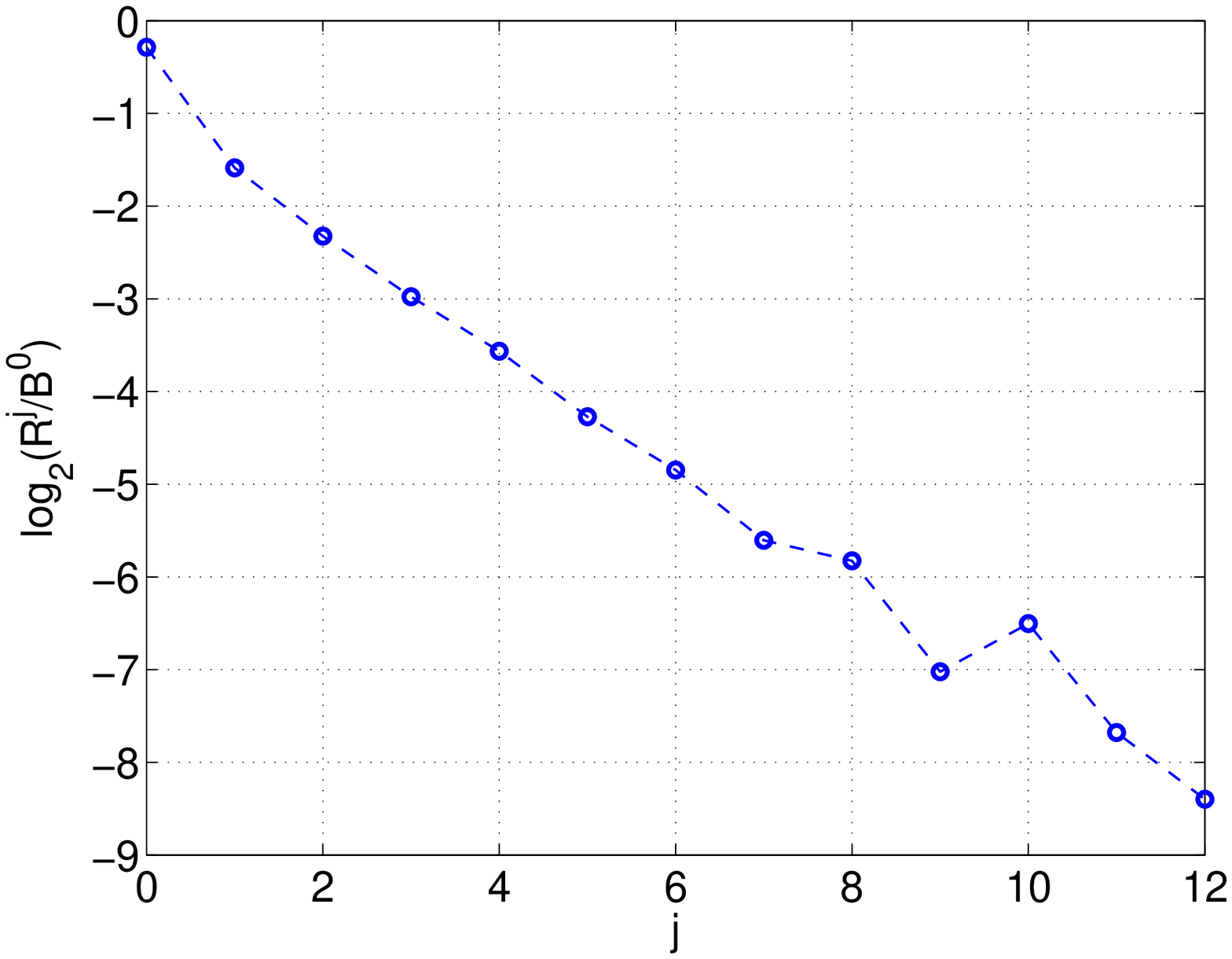}
\caption{For an $N=2^{16}$ random  normally-distributed time series, with variance $\sigma^2=4$,
 as a function of $j$: (a) log$_2$ of the mean number of extrema, normalized
to~$N$;
 (b) log$_2$  of the mean distance between the extrema of the baselines;
  (c) log$_2(\mbox{mean}\|B^j\|_2/\mbox{mean}\|Z\|_2)$; (d) log$_2(\mbox{mean}\|R^j\|_2/\mbox{mean}\|Z\|_2)$. 
 } 
\label{fg.allmaxU}
\end{figure}

Figure \ref{fg.BRscaling} illustrates a typical ITD decomposition
of a noisy signal; Figures \ref{fg.spectral},~\ref{fg.allmaxN}, and~\ref{fg.allmaxU} depict
empirical scaling properties that will be studied quantitatively in
Section \ref{diffusion}. 
Panel (\ref{fg.spectral}a) shows the spectum of the energy of (\ref{zeq}). Panels
(\ref{fg.spectral}b) and (\ref{fg.spectral}c) show the
normalized enegy spectrum of $B$ and $R$, respectively. 
We can see how the energy is shared between baseline and rotation:
the ratio  $\|B\|_2/\|Z\|_2$ 
is about $0.37$, and
the ratio of $\|R\|_2/\|Z\|_2$ is about $1.77$. (Subscript~$2$ denotes the
$\ell^2 $ norm.)

Next, we estimate the rate at which the wavelength of the baselines in our all-extrema
signal increases as the high-frequency components $R^j$ are removed.
We measure this by computing the ratio of the spacings between
extrema from one stage to the next, averaged over an ensemble of 
decompositions of  random all-extrema signals of the same length and statistical distribution.
Experiments of this kind were done for the EMD in
\cite{wufilterbank}, \cite{flandrinfilterbanks};
because of the very regular
scaling behavior, the EMD could be interpreted
as a filter bank. We found that the ITD has similar
scaling universality, and offer a partial
analytical explanation in the next section.

The slopes of the lines in panels (\ref{fg.allmaxN}a) and 
(\ref{fg.allmaxN}b) and the data in Table~\ref{table1}
show that the spacing of the extrema of the $B^j$ increases by a factor~$2.6$,
and the number of extrema drops by a factor~$.4$, as~$j$ increases.
The energy ratio $\|B^j\|_2/\|Z\|_2$ drops by about 0.4 for~$j=1$, and by 
about $ 0.63$ for $j>1$.  As would be expected, 
the number~$D$ of levels required for a full decomposition increases with~$N$.
Table \ref{table1} summarizes the data
for the $N=512$ case in Figure \ref{fg.allmaxN}, up to 
level $j=6$. 
\begin{table}[ht]
\centering
 \caption{Analysis of the average of the first six levels of an ITD decomposition of  all-extrema signals,
 length $N=512$,
  with $z_i$ drawn from ${\cal N}(0,4)$. See Figure \ref{fg.allmaxN}. Average results from 50,000 experiments (with $B^0=Z$ of the same length and
 statistical distribution).  NE denotes the 
 average number of extrema, normalized to $N$.  DE refers to the average distance between extrema. In the last column, subscript~$2$ denotes the $\ell_2$ norm.} 
 \label{table1}
 \begin{tabular}{|c||c|c|c|c|}
 \hline
j & NE & DE& NE$\times$DE & $\mbox{mean} \|B^j\|_2$/$\mbox{mean}\|B^0\|_2$\\
\hline
0 & 1.000  &1.00  &1.000   &1.00  \\
 1  & 0.422 &2.38 & 1.00  & 0.37\\
 2  & 0.166 &  6.10 & 1.01 &0.21 \\
 3  & 0.065 &15.9 &  1.04  &0.14  \\
 4  & 0.026  & 42.0  &1.1  &  0.09 \\
 5  & 0.017  & 112 & 1.3  & 0.06\\
 6  & 0.006 & 290 & 1.8  &  0.04 \\
   \hline
  \end{tabular}
\end{table}
The trends shown in the table and the figures were very stable to changes in the variance of the original signal changes in the outcomes of the order of tenths of a percent for a range of variances between 1 and 20.
We also tested an all-extrema 
time series with the $z_i$ drawn from a uniform distribution,
and found that  the scaling factors are close to those of the normal case reported above.

A decomposition of a signal that consists of   $2^{16}$ normal variates 
drawn from 
$\mathcal{N}(0,4)$ (discrete white noise) yields the results portrayed in Figure \ref{fg.allmaxU}. 

 There were $13$ baselines (approximately $\log_2 2^{16}$).
From the slopes of the lines in
panels (\ref{fg.allmaxU}a) and (\ref{fg.allmaxU}c) and the corresponding data
(not shown) we estimate
that the number of extrema again drops by the factor $0.4$,
the distance between extrema increases by the factor $2.55$, 
and the normalized $\ell^2$ of the baselines decreases by about $0.61$. 
The analytical model developed in Section~3 yields a value of $0.55$.
The scaling
pattern deteriorates as the baselines and rotations flatten.

\section{Intrinsic Time-scale Decomposition of Random Signals: Universality}
\label{diffusion}

In this section, we will attempt to understand the scaling laws from the section~\ref{sec:random} that were obtained numerically for  ITD applied to random signals.  We first propose a surrogate model for the baselines of a random ITD signal using the scaling/translation symmetries of the ITD process and intuition gained from numerical experiments. We then validate the surrogate model by comparing predictions of the surrogate model with the ITD of random Gaussian signals. This comparison also suggests ways to improve the surrogate model. Finally, we analyze one step of the ITD process applied to the surrogate baselines, and this analysis helps explain the observed self-similarity of the ITD baselines for random signals, and also provides estimates for the decay rates for the $L^2$ norm and the number of extrema in the baselines.

\subsection{Surrogate model for the baselines}

Associated with the ITD at level $j$, we define the set $S^j = \{\tau^j_1,\tau^j_2,\ldots,\tau^j_{m^j}\}$ of cardinality $m^j = |S^j|$, the location of the extrema in $B^j$, and the vector $b^j \in \mathbb{R}^{m^j}$, the values of the baseline $B^j$ at the extrema. We  denote by $\mathcal{E}$ the operator that extracts the locations and values of the extrema of an arbitrary  time series, so in particular,
$\mathcal{E}[B^j] = \{S^j,b^j\}$. $\mathcal{E}$ is a nonlinear but homogeneous operator i.e. $\mathcal{E}[ c B^j] = \{S^j,c b^j\}$ for any constant $c \neq 0$.

To determine $\{S^{j+1},b^{j+1}\}$ we do not need to know the entire baseline $B^j$; it suffices to know $\{S^j,b^j\}$ (See Eq.~(\ref{bk})). The ITD procedure therefore gives a reduced dynamics on
the pairs
$\{S^j,b^j\} = \mathcal{E}(B^j)$. The operator $\mathcal{E}$ is not one-to-one, and hence not invertible. In order to compare the reduced dynamics on $\{S^j,b^j\}$ with the ``full" ITD baselines $B^j$, we define a {\it surrogate baseline} $\tilde{B}^j$ by
$
\tilde{B}^j = \sum_{k = 1}^{m^j} b^j_k e^j_k
$
where $e^j_k$ is a piecewise linear function (time-series) which is~$1$
at location $S^j_k$ and~$0$
 on every other $S^j_\ell$. Then
 $\mathcal{E}(\tilde{B}^j) = \mathcal{E}(B^j) = (S^j,b^j)$ and since $B^j$ and $\tilde{B}^j$ are both monotone  between their (common) extrema,
 we expect  $\tilde{B}^j$ to be a good approximation to $B^j$. The surrogate baseline is a ``rough" analog of the
 IMFs in the EMD method; in that construction, the modes
arise from cubic spline interpolations of the maxima and
minima in the signal  \cite{huangemdfirst}.

\begin{figure}[htbp]
\centering

(a)\includegraphics[width=2.1in,height=0.7in]{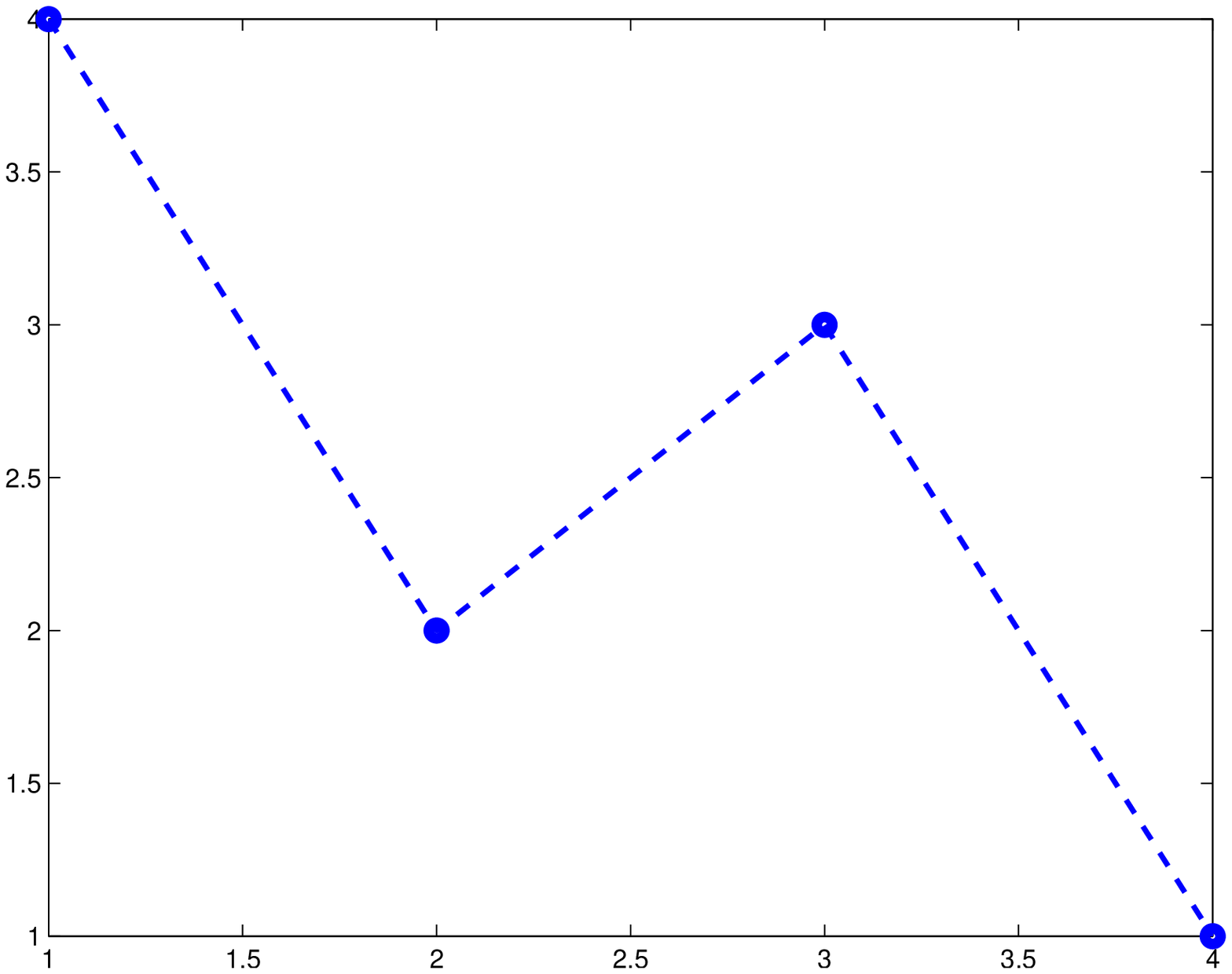}
(b)\includegraphics[width=2.1in,height=0.7in]{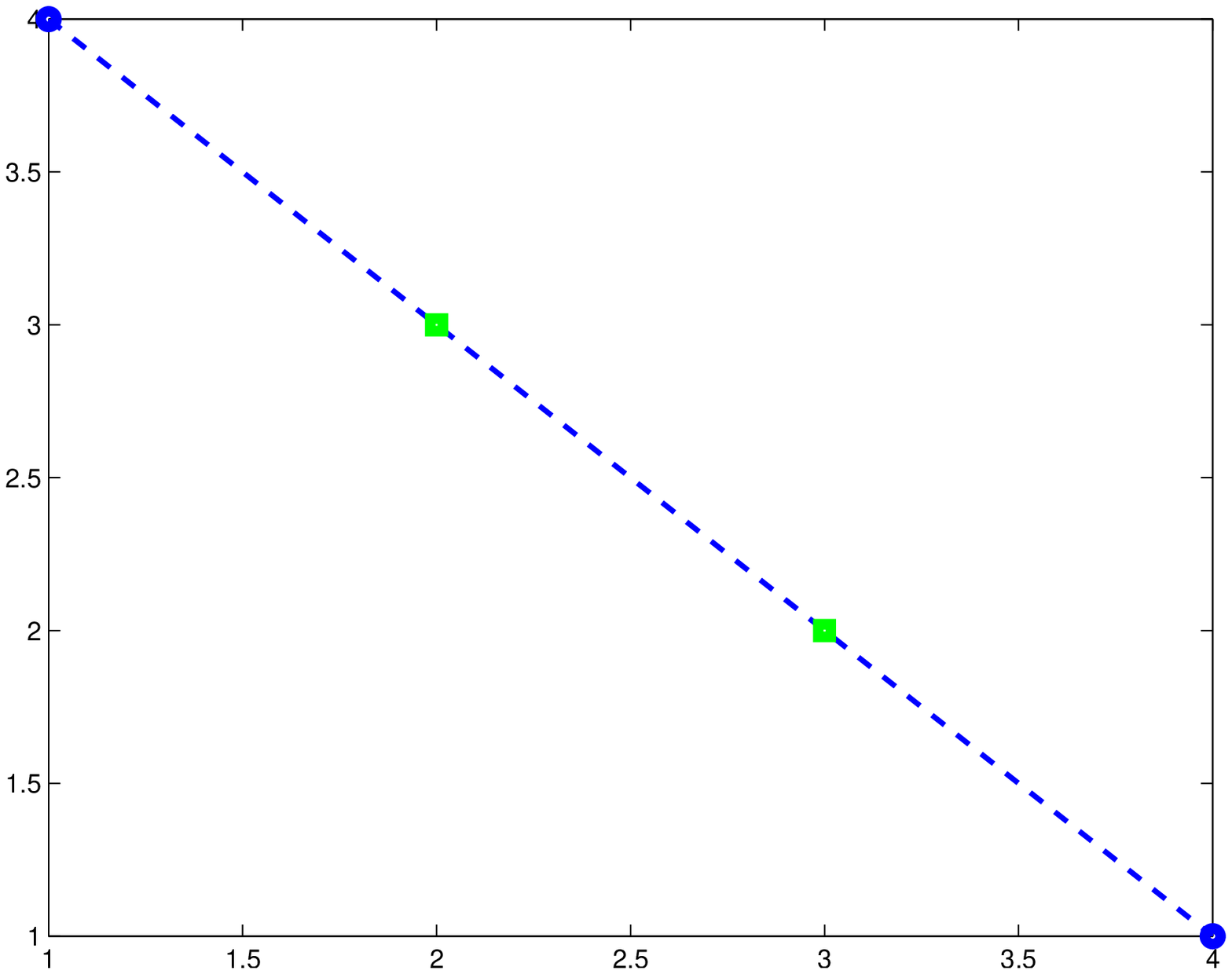}

(c)\includegraphics[width=2.1in,height=0.7in]{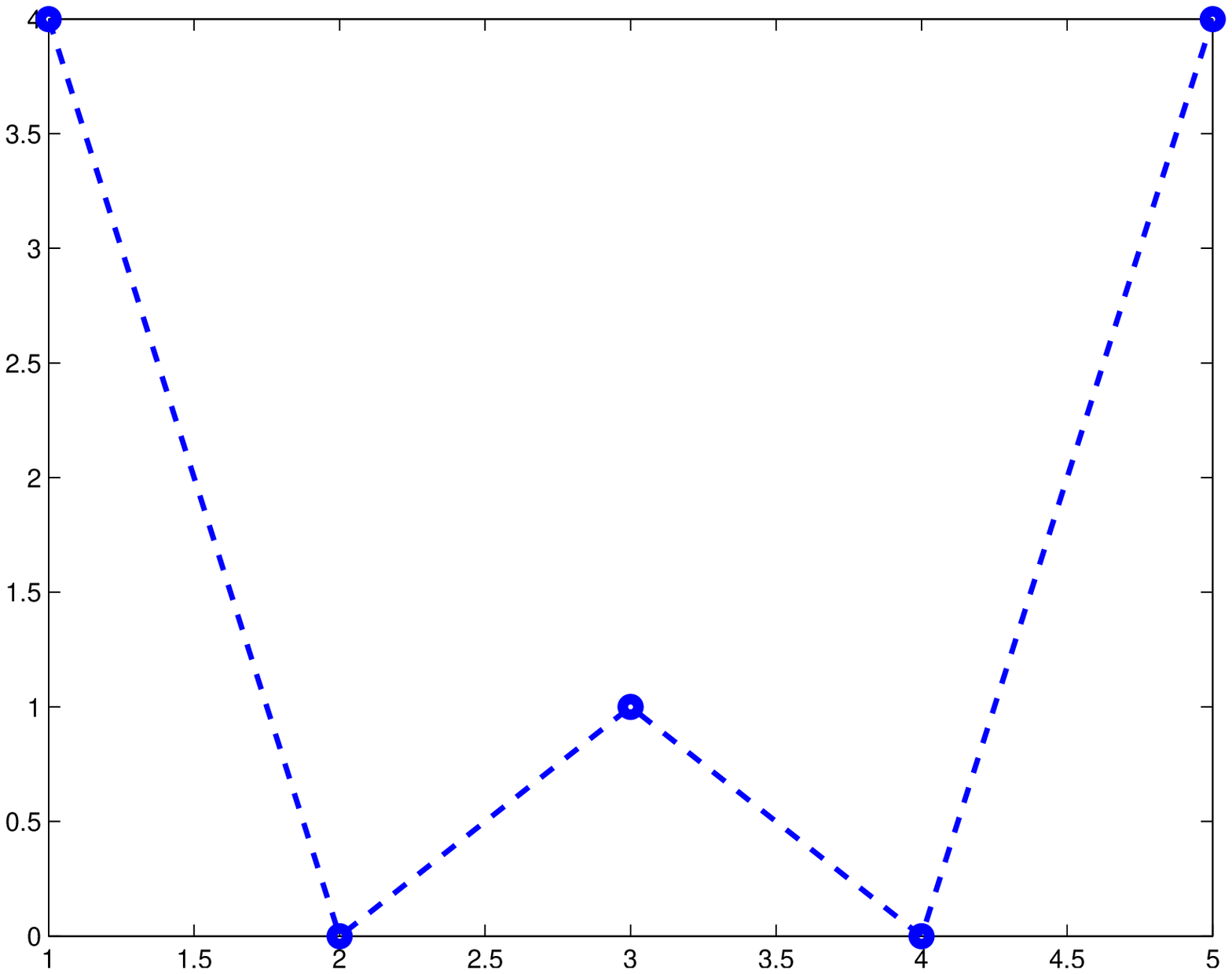}
(d)\includegraphics[width=2.1in,height=0.7in]{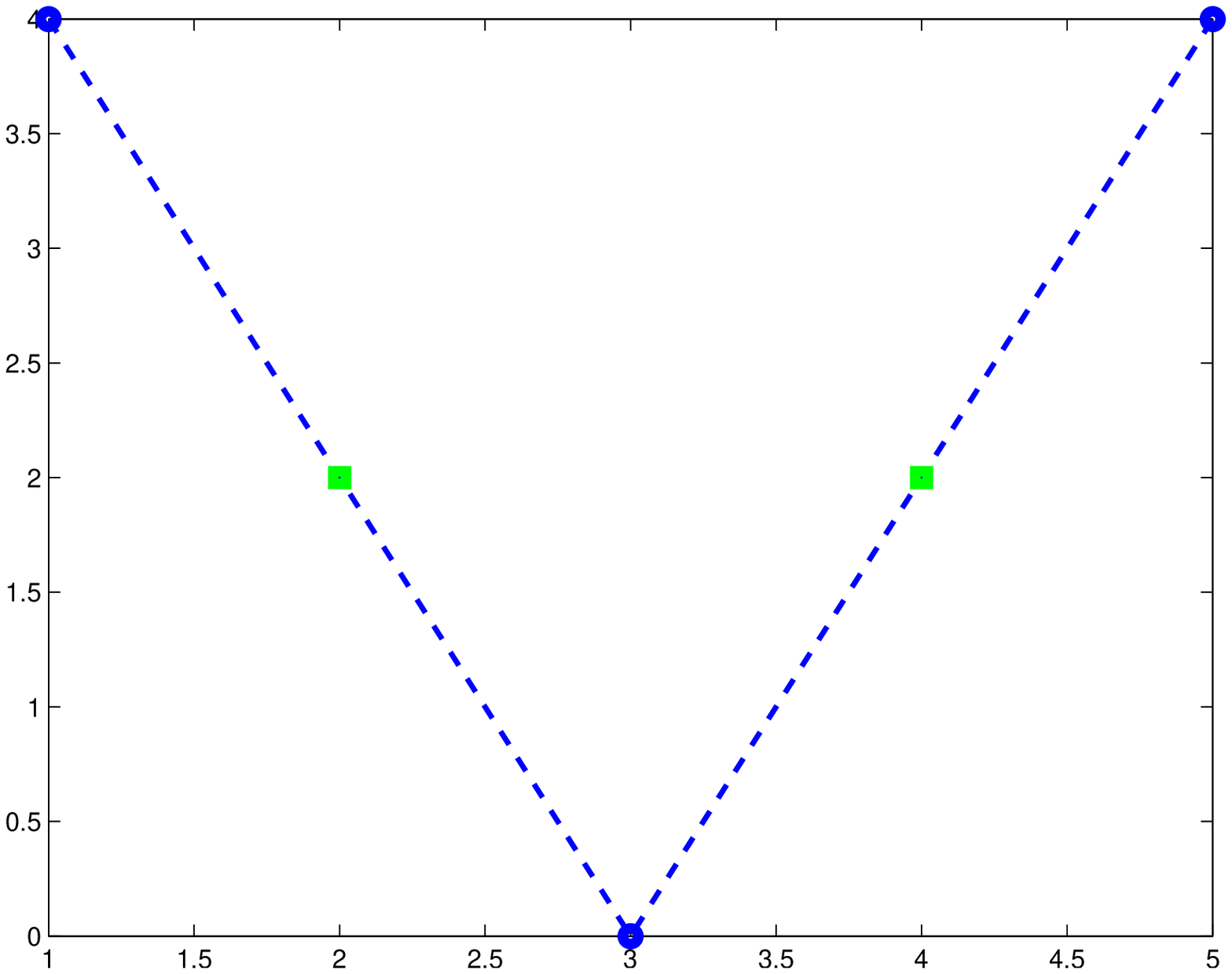}

   \caption{
   The dark circles represent extrema, and the light squares are points which go from being extrema at level $j$ to not being extrema at level $j+1$.
 }
   \label{fg.bifurcate}
   \end{figure}

 Certain extrema in $S^j$ ``disappear'' in $S^{j+1}$, so that
 the extrema at level $j+1$ satisfy $S^{j+1} \subseteq S^j$ .
 There are two types of processes which decrease the number of extrema. These are illustrated in Fig.~\ref{fg.bifurcate}. In the top panel,
neighboring extrema
flip their relative positions; in dynamical systems language, this is a saddle-node
bifurcation. In the bottom panel
an extremum changes type and its two neighbors disappear; this is a pitchfork
bifurcation.
We ran the
  ITD process with $2^{16}$ initial points and computed the frequencies
 of occurrence of the two bifurcation types. After an initial transient (corresponding to $j =1$ and 2) the saddle-node bifurcation occurs
with probability $\gamma \approx 0.58$, and the pitchfork bifurcation with probability
$\beta \approx 0.21$, and these probabilities are independent of the level $j$. Also,
$(1 - \gamma - \beta) \approx 0.21 \approx \beta $, which corresponds to the probability of no (local) change in the nature of the extremum. This gives the ({\em a priori} unexpected) conclusion that every local maximum at level $j$ remains a maximum or becomes a local minimum with roughly equal probabilities $\beta$ at level $j+1$. Further, these probabilities are independent of $j$.

Our numerical experiments suggest that after a few iterations, usually one or two,
the extrema disappear independently of their neighbors. At that stage,
the sets $S^j$ evolve by an independent random decimation process, so the ``lifetime" for any given point $x \in S^1$ as an extremum, i.e, the maximal $j$ such that $x \in S^j$, has a geometric distribution with
parameter $\gamma$ (the probability of losing an extremum via the pitchfork
bifurcation).
The probability that the lifetime equals $j$ is $(1-\gamma)^{j-1} \gamma$.

\begin{figure}[htbp] 
   \centering
   \includegraphics[scale=.3]{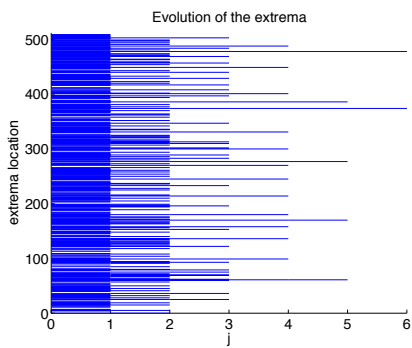}
   \caption{A graphical representation of the sets $S^j$ for 6 levels of an ITD starting with a random time series consisting of 512 normal variates.}
   \label{fig:extrema}
\end{figure}

The initial distribution of the inter-extremal spacings is given by the chosen initial conditions. E.g., the distribution is concentrated at $l=1$ for the all-extremum signal $Z(i)$ in (\ref{zeq}).  Evolution by independent random decimation at each extremum implies that each site at level $j$ the inter-extremal spacings~$l_k$ are a sum of a random number $n_k^j$ of ``initial" separations, where $n^j_k$ is drawn from a geometric distribution. After an initial transient, $n^j_k \gg 1$, so the law of large numbers will imply that $l_k \approx n^j_k E[l^0_k]$  were $E[l^0_k]$ is the average spacing between extrema in the initial condition. Consequently, $l_k$ is approximately geometrically distributed with a $j$-dependent mean denoted by~$\lambda_k$. This agrees qualitatively with numerical simulations of the ITD with random gaussian initial conditions, as shown in Fig.\ref{fig:extrema}. Numerical experiments also suggest that the initial transient is short, typically $j=1$ or $2$ ITD steps.

The ITD algorithm Eq.~(\ref{bk}) which computes the extrema at level $j+1$ can be written as
$
\{S^{j+1},b^{j+1}\} = \mathcal{E}[(I + M^j) b^{j}],
$
where $I$ is the identity matrix and
$M^j$ is a matrix whose rows sum to zero; when we are only interested in baseline
extraction, we omit the symbol $S^{j+1}$.
$(I+M^j)$ is thus a stochastic matrix, and the entries  of $M^j$ are determined by  $S^j$ via (\ref{bk}).
In particular, for any vector $b$, if $v = M^j b$, then
\begin{equation}
v_{k} = \frac{1}{4}(b_{k-1}-2 b_k + b_{k+1}) + \frac{q^j_k}{4} (b_{k+1}-b_{k-1})
\label{eq:diffusion}
\end{equation}
where $q^j_k \in (-1,1)$ is given by
\begin{equation}
q^j_k = \frac{(\tau^j_{k}-\tau^j_{k-1})-(\tau^j_{k+1}-\tau^j_{k})}{(\tau^j_{k}-\tau^j_{k-1})+(\tau^j_{k+1}-\tau^j_{k})} = \frac{ 2\tau^j_{k}-\tau^j_{k-1} - \tau^j_{k+1}}{\tau^j_{k+1}-\tau^j_{k-1}}.
\label{eqp}
\end{equation}
(See Eq. (\ref{alt3})).
The parameter $q_k^j$ measures the asymmetry in the distances of the knot $\tau_k^j$
from the neighboring extrema.
If the vector $b$ is obtained by sampling a smooth function $B(x)$, then $M^j b$ can be interpreted as sampling $\frac{1}{4}B''(x) + \frac{1}{2}q^j(x) B'(x).$
We can thus interpret $(I+M^j)b^j$  as the numerical solution at one time step of the forward-time, center-difference approximation to the solution to
$$
\frac{\partial}{\partial t} B = \frac{1}{4}\frac{\partial^2}{\partial x^2} B + \frac{1}{2}q^j(x) \frac{\partial}{\partial x} B = \frac{1}{4 w^j(x)} \frac{\partial}{\partial x}\left[ w^j(x) \frac{\partial B}{\partial x}\right],
$$
where $w^j(x) = \exp\left[2 \int_0^x q^j(t) dt \right]$.

In what follows, we will be assuming periodic boundary conditions, so there are as many local minima as maxima, and the cardinality of $S^j$ is always even.   If the underlying time signal is mean zero and stationary, then the expected value of a maximum is the negative of the expected value of a minimum. We now make two approximations to obtain a form for $b^j_k$.
First,
we assume that each maximum or minimum is  a random
Gaussian perturbation of the expectation.
Second, we postulate that the values of the maxima and minima are independent
random variables (for a test of this assumption, see below, and Figure~\ref{fig:corr} (b)).
It now follows that
$$
b^j_k \approx \mu^j (-1)^k + \alpha^j n_k
$$
where $\mu^j$ is the mean value of the maxima (or the negative of the minima) in $B^j$, the $n_k$ are independent normal variates and $\alpha_j^2$ is the variance of the maxima (or also the minima). The fact that $\alpha_j$ only depends on $j$ and not on $\tau^j_k$ is a consequence of the underlying random process being stationary.

We can test this {\it ansatz} numerically by computing the auto-correlation
$
R^j(l) := E[b^j_k b^j_{k+l}] = (\mu^j)^2 (-1)^l + (\alpha^j)^2 \delta_l
$
where $\delta$ is the Kronecker delta.  Fig.~\ref{fig:corr}a depicts the average over 100 runs of the normalized autocorrelation $R(l)/R(0)$ for lags $0 \leq l \leq 31$ for the first 6 levels of the ITD (the six curves are superimposed). Note that the auto-correlation is for the signal $b^j$ at level $j$ which consists of only the extremal values (the signal sampled at $\tau^j_k$ and then exhibited as a function of $k$), and not the full baseline $B^j$.  In each run, the initial time series has $2^{16}$ i.i.d normal variates. $R(l)/R(0) =1$ for $l = 0$ (zero lag) and otherwise $|R(l)/R(0)| \leq 1$. As one would expect, there is a high frequency oscillation in the auto-correlation corresponding to the alternation between maxima and minima. We can  remove this oscillation by considering the absolute value of the autocorrelation
$
|R(l)| = (\alpha^j)^2 \delta_l + (\mu^j)^2.
$
The assumed {\em ansatz} for $b^j_k$ thus predicts that $|R(l)|/R(0)$ should be a constant, less than 1, for all $l \neq 0$.

Figure~\ref{fig:corr} (b) shows $|R(l)/R(0)|$ for different levels $j$. After an initial transient, the normalized correlations collapse on to a single universal curve for  $j\geq 3$. Further, this universal curve is well described by a single, $j$ independent, constant, except for persistent deviations at $l=1$ and $l=2$. This implies there is a universal self-similar description of $b^j_k$ for large $j$, and there is indeed a short range correlation between the extrema (nearest neighbor $l=1$ and next nearest neighbor $l=2$). Our assumption, that $b^j_k \approx \mu^j (-1)^k + \alpha^j n_k$ where the $n_k$ are independent, can likely
be  improved by accounting for this correlations between the values of the extrema.
\begin{figure}[htbp] 
   \centering
(a)\includegraphics[width=2.2in,height=1in]{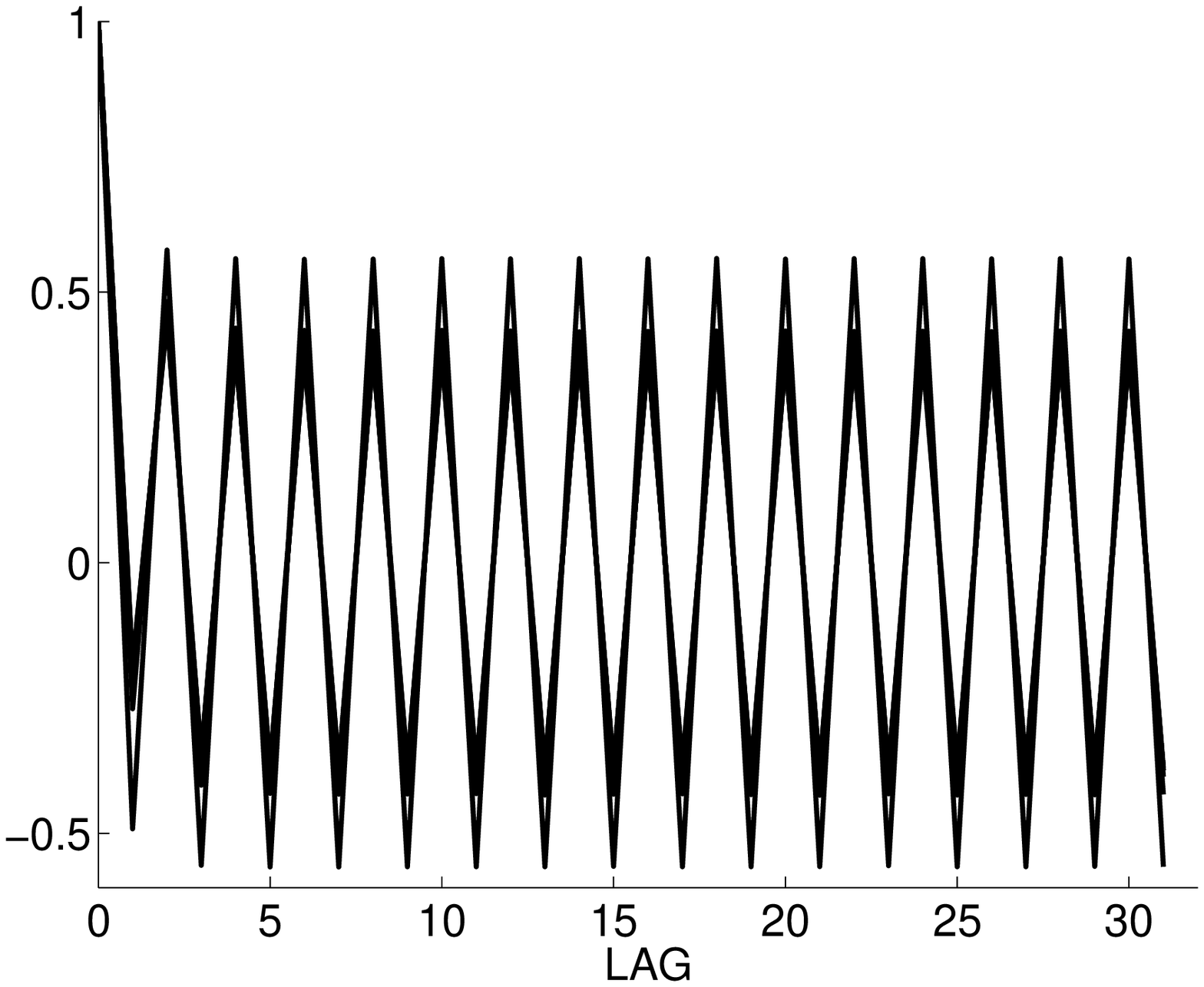}
(b)\includegraphics[width=2.2in,height=1in]{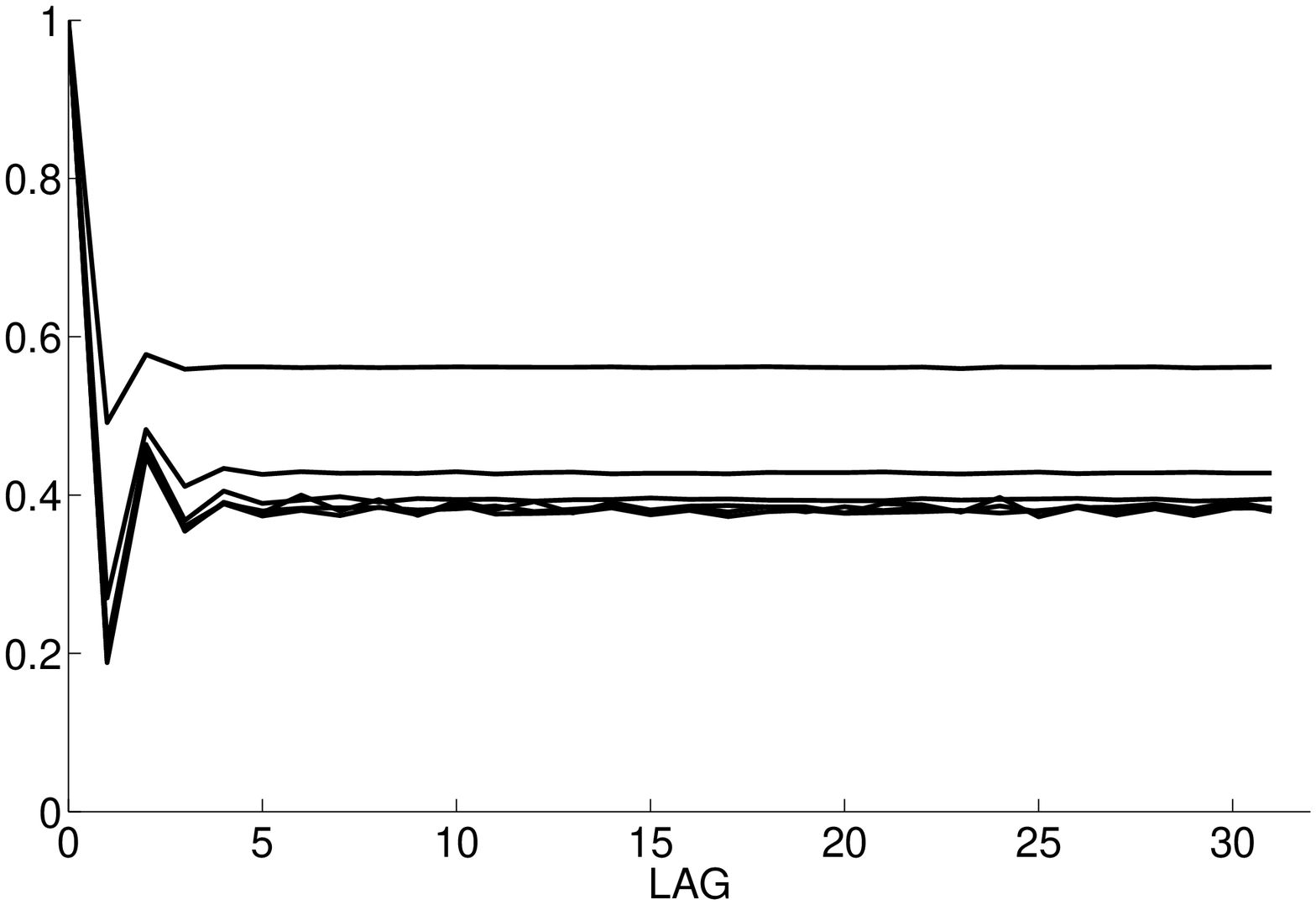}
   \caption{(a) The averaged auto-correlation at the first 6 levels of the ITD normalized by the $\ell_2$-norm. (b) The absolute value of the averaged and normalized autocorrelation. }
   \label{fig:corr}
\end{figure}

\subsection{Analysis: Universality and decay rates}

Given this approximate description of the signal $b^j$,
we can now compute the signal $b^{j+1}$ and also
the surrogate baseline $\tilde{B}^j$, and thus study the evolution of the baselines as a function of the index $j$ of the ITD.
Since $(I+M^j) (-1)^k = 0$, the mean periodic oscillation between the maxima and the minima is in the null space of the matrix $(1+M^j)$.  Therefore $(1+M^j)b^j = \alpha^j(1+M^j)n_k$ and $b^{j+1} =
\alpha^j \mathcal{E}((I+M^j) \mathbf{n})$,
where $\mathbf{n} = n_k$ is a vector of independent normal variates. This motivates the consideration of the signal $\mathcal{E}((I+M^j) \mathbf{n})$.
If $x_1, x_2$ and $x_3$ are consecutive entries of the vector $(I+M^j) \mathbf{n}$, we have
 \[
 \mathbf{x} = \left( \begin{array}{c} x_1 \\ x_2 \\ x_3\end{array} \right) \approx \frac{1}{4}\left( \begin{array}{ccccc} 1-q_1 & 2 & 1+q_1 & 0 & 0 \\
 0 & 1-q_2 & 2 & 1+q_2 & 0 \\  0 & 0 & 1-q_3 & 2 & 1+q_3 \end{array}\right) \left(\begin{array}{c}   n_1 \\ n_2 \\ n_3 \\n_4 \\ n_5 \end{array} \right) \equiv A \mathbf{n}.
 \]
where $n_1,n_2,n_3,n_4$ and $n_5$ are independent normal variates
and $q_k$ is defined in (\ref{eqp}).
For every given realization of $q_1,q_2$ and $q_3$, the entries
$x_1,x_2$ and $x_3$ are jointly Gaussian with mean zero and covariance
\[
\Sigma(q_1,q_2,q_3) = AA^T=
\frac{1}{16} \left(\begin{array}{ccc} 6 + 2 q_1^2 & 4 + 2 q_1 - 2q_2 & (1+q_1)(1-q_3)  \\
  4 + 2 q_1 - 2q_2 & 6 + 2 q_2^2  &  4 + 2 q_2 - 2q_3  \\ (1+q_1)(1-q_3) & 4 + 2 q_2 - 2q_3 &  6 + 2 q_3^2 \end{array} \right)
 \]
 The conditional joint density of $x_1,x_2$ and $x_3$ is given by
 \begin{equation}
 p(x_1,x_2,x_3|q_1,q_2,q_3) = \frac{1}{\sqrt{8 \pi^3 \mathrm{Det}(\Sigma(q_1,q_2,q_3))}} \exp\left[-\frac{1}{2} \mathbf{x}^T \Sigma(q_1,q_2,q_3)^{-1} \mathbf{x}\right].
 \label{cond-gauss}
 \end{equation}
To proceed further, we now compute the joint density of $q_1,q_2$ and $q_3$. For this we need the (as yet unknown) distribution of the inter-extremal separations $l_k = \tau_{k+1} - \tau_k$. A typical realization of the sets $S^j = \{\tau^j_1,\tau^j_2,\ldots\}$ is shown in Fig.~\ref{fig:extrema}. As we argued earlier, at every level $j$ the inter-extremal  separations $l_k$ have a geometric distribution, with a parameter that depends on $j$. If the number of nodes is large, then we can ignore the discrete nature of the underlying sets $S^j$ and consider instead the  exponential distribution which is the continuous analog of the discrete distribution. The probability density of the inter-extremal separation is then given by $p^j(l) = \lambda_j^{-1}\exp(-l/\lambda_j)$ where $\lambda_j$ is the mean inter-extremal spacing at ITD level $j$ (See figure~\ref{fig:extrema}).

 The variables
  $q^j_1$ and $q^j_2$ are defined by ratios of $l_1,l_2$ and $l_3$ in Eq.~(\ref{eqp}), so their distribution does not depend on the parameter $\lambda$  defining the mean of the exponential distribution. Alternatively, we are free to pick our unit for length for $l_1,l_2$ and $l_3$ as the mean of the exponential distribution for $l_k$, and this does not affect $q_k$ which are
  non-dimensional. Without loss of generality, we can thus assume the mean inter-extremal spacing is 1. The probability $P(y,z) =\mbox{Prob}((q_1 > y) \mbox{ and } (q_2 < z))$ is the probability of the event $l_1 > \frac{1+y}{1-y} l_2$ and $l_3 > \frac{1-z}{1+z} l_2$,
$$
  P(y,z) = \int_0^\infty dl_2 \int_{ \frac{1-z}{1+z} l_2}^\infty d l_3 \int_{\frac{1+y}{1-y} l_2}^\infty dl_1 e^{-(l_1+l_2+l_3)} = \frac{(1-y)(1+z)}{3-y+z + yz}.
  $$
The joint density of $q_1$ and $q_2$ is given by
$$
\rho(q_1,q_2) = \left. -\frac{\partial^2}{\partial y \partial z} P(y,z)\right|_{y = q_1,z=q_2} = \frac{8(1-q_1)(1+q_2)}{(3-q_1+q_2 +q_1 q_2)^3}.
$$
We can also compute the marginal distribution of $q_2$ by
$$
\mbox{Prob}(q_2 > z) = P(-1,z) = \frac{1+z}{2},
$$
showing that $q_2$ is uniformly distributed in $(-1,1)$. This yields the conditional density
$$
\rho(q_1|q_2 = z) = \frac{16(1-q_1)(1+z)}{(3-q_1+z +q_1 z)^3}.
$$
Equation (\ref{eqp}) shows that there are no common intervals  $l_k$
in the definition of $q_3$ and $q_1$,
and by translation invariance, the joint distribution of $q_2$ and $q_3$ is identical in form to the computed joint distribution of $q_1$ and $q_2$. The joint density of $q_1,q_2$ and $q_3$ is therefore
\begin{equation}
\rho(q_1,q_2,q_3) = \frac{128(1-q_1)(1+q_2)(1-q_2)(1+q_3)}{(3-q_1+q_2 +q_1 q_2)^3 (3-q_2+q_3 +q_2 q_3)^3}.
\label{joint}
\end{equation}
The joint density for $(x_1,x_2,x_3,q_1,q_2,q_3)$ is the product of the densities in Eqs.~(\ref{cond-gauss})~and~(\ref{joint}). An important observation is that the joint density is {\em independent of the level $j$} of the ITD so we {\em should expect} self-similar behavior in the ITD decomposition of a random signal.

We will define an (approximate) marginal distribution on $x_1,x_2$ and $x_3$ by positing that this distribution is still jointly Gaussian. We can compute the covariance of this distribution as
\begin{eqnarray*}
\Sigma & = \int_{-1}^1 dq_2 \int_{-1}^1 dq_1 \int_{-1}^1 dq_3 \Sigma(q_1,q_2,q_3) \rho(q_1,q_2,q_3)  \\
& = \left(\begin{array}{ccc} 0.42\ldots & 0.25 & 0.058\ldots \\ 0.25 & 0.42\ldots & 0.25 \\ 0.058\ldots & 0.25 & 0.42\ldots \end{array} \right).
\end{eqnarray*}
We now obtain the joint density of $x_1,x_2$ and $x_3$ by integration,
 $$
 p (x_1,x_2,x_3) \approx \frac{1}{\sqrt{8 \pi^3 \mathrm{Det}(\Sigma)}} \exp\left[-\frac{1}{2} \mathbf{x}^T \Sigma^{-1} \mathbf{x}\right].
 $$
 The probability $\beta$ that $x_2$ is a local maximum is the probability of the event $x_1 < x_2$ and $x_2 > x_3$, i.e.
 $$
 \beta = \int_{-\infty}^{\infty} dx_3\int^{\infty}_{x_3} dx_2 \int^{x_2}_{-\infty} dx_1 p(x_1,x_2,x_3) \approx 0.24.
 $$
  The probability that a given site is a local minimum is also $\beta$ since
  $$
   \int_{-\infty}^{\infty} dx_3\int^{\infty}_{x_3} dx_2 \int^{x_2}_{-\infty} dx_1 p(x_1,x_2,x_3)  =  \int_{-\infty}^{\infty} dx_3\int_{-\infty}^{x_3} dx_2 \int_{x_2}^{\infty} dx_1 p(x_1,x_2,x_3)
   $$
by the symmetry of $p$, thus  explaining the observation that an extremum in ITD level $j+1$ was equally likely to be a maximum or a minimum independent of its type at level $j$. The decay rate for the number of extrema is given by
 \begin{equation} \label{maxdecay}
 m^{j+1} \approx 2 \beta m^j, \quad    \mbox{which implies that} \quad m^j  \approx (0.48)^j m^0,
 \end{equation}
which is in approximate agreement with the numerically determined decay rate of $0.4$ for the number of extrema is ITD for a random i.i.d Gaussian signal (figure~\ref{fg.allmaxU}).

 We can also compute the mean and the variance of the distribution of the maxima of $(I+M^j) \mathbf{n}$ by the conditional expectations
 $$
 \mu = E[x_2| x_2> \max(x_1,x_3)] = \frac{1}{\beta} \int_{-\infty}^{\infty} dx_3\int^{\infty}_{x_3} dx_2 \int^{x_2}_{-\infty} dx_1 \, x_2\,  p(x_1,x_2,x_3) = 0.48
 $$
  and
 \begin{eqnarray}
 \alpha^2 & = & E[x_2^2| x_2> \max(x_1,x_3)] - \mu^2 \nonumber \\
 & = & \frac{1}{\beta} \int_{-\infty}^{\infty} dx_3 \int^{\infty}_{x_3} dx_2 \int^{x_2}_{-\infty} dx_1 \, x_2^2 \, p(x_1,x_2,x_3) - \mu^2  \nonumber \\
 & = & 0.30 \nonumber
\end{eqnarray}
From this, we obtain
$
\mathcal{E}((I+M^{j-1}) \mathbf{n}) = (S^{j},b^{j})$ where $b^{j} \approx \mu (-1)^k + \alpha \mathbf{n}' = 0.48 \times (-1)^k + 0.55 \mathbf{n}',
$
where $\mathbf{n}'$ is a vector of $|S^{j}|$ i.i.d normal variates. If $\tilde{B}^{j}$ is the piecewise linear interpolating function defined by the extremal values $b^j_k$ on the set $S^{j} = \{\tau_1^{j},\tau_2^j,\ldots,\tau_{m_j}^j\}$, then a direct calculation shows that
$$
\sum_i |\tilde{B}^j_i|^2 \approx \sum_i E\left[|\tilde{B}^j_i|^2\right] \approx \sum_{k=1}^{m_j} \frac{\mu^2 + 2 \alpha^2}{3} l_k =  \frac{L}{3}(\mu^2 + 2 \alpha^2),
$$
where $L = \sum l_k$ is the total number of data points in the time series, and we are assuming that $|S^j| = m_j \gg 1$ so we are justified in replacing the (random) sum by its expected value. The second approximation is replacing the sum by the corresponding integral which is valid for $L \gg 1$. Observe that the $\ell^2$ norm of the surrogate baseline $\tilde{B}^j$ only depends on the two numbers $\mu$ and $\alpha$, and {\em not on the set $S^j$!}.

Since $\mathcal{E}((I+M^{j-1}) \mathbf{n}) = (S^{j},b^{j})$ and $\mathcal{E}(\alpha (I+M^{j}) \mathbf{n'} ) = (S^{j+1},b^{j+1})$, it immediately follows that $\|\tilde{B}^{j+1}\|/\|\tilde{B}^j\| = \alpha = 0.55$. This also gives a decay rate for the $\ell^2$ norm equal to $\alpha
 \approx 0.55$ in comparison to the numerically obtained figure of $0.61$ (See figure~\ref{fg.allmaxU}).  This estimate, as well as the decay rate of
the number of maxima in (\ref{maxdecay}), both exceed the empirical parameters by
about 17\%.  We saw that extrema disappear through
nearest neighbor interactions, Figure~\ref{fg.bifurcate}, and it is plausible that
the nearest-neighbor correlations seen in Figure~\ref{fig:corr} cause extrema to
persist longer than is predicted when independence is postulated.

Finally, we observe that $\mu^2/(\mu^2 + \alpha^2) \approx
0.45$ in good agreement with $|R(l)|/R(0)$ for $l \neq 0$ in Fig.~\ref{fig:corr}(b). We believe
that this type of analysis can be extended to EMD and it is an
interesting question whether this will explain the observed
self-similar behavior in EMD for random signals \cite{wufilterbank,flandrinfilterbanks}.

\section{End effects} \label{endconditions}

The results of the ITD decomposition of a signal may depend strongly on the
boundary conditions.
Consider the time series 
$
f(t_i) =  0.5 e^{t_i}  +a \cos(10 t_i), \quad t_i =\{0:0.01:2 \pi\},$ where $a=10$ or $a=50$.
Variation of the parameter~$a$ will cause significant changes
in the baselines at the right endpoint.
\begin{figure}[hbt]
\centering
(a)\includegraphics[width=6.1cm,height=1.5in]{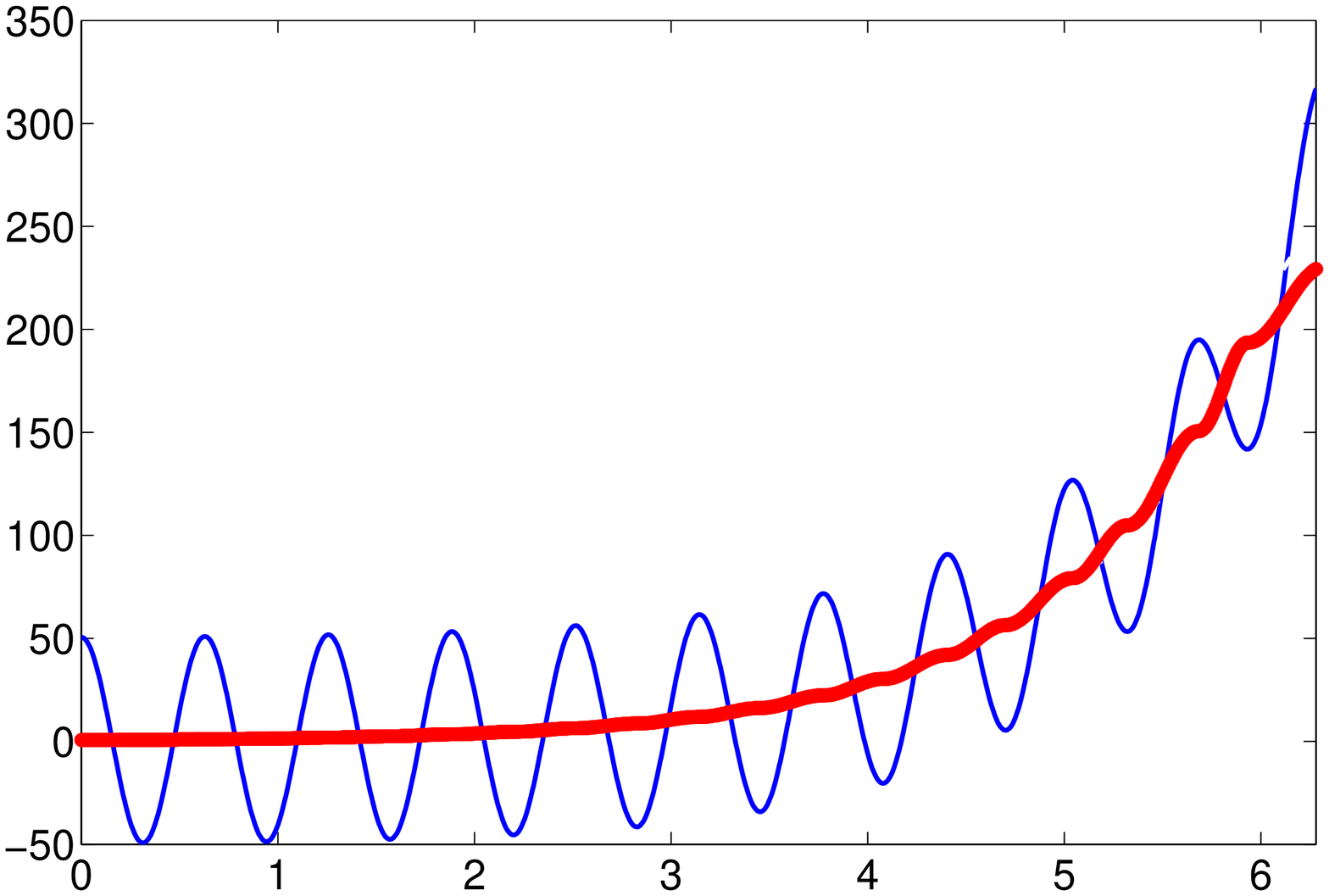}
(b)\includegraphics[width=6.1cm,height=1.5in]{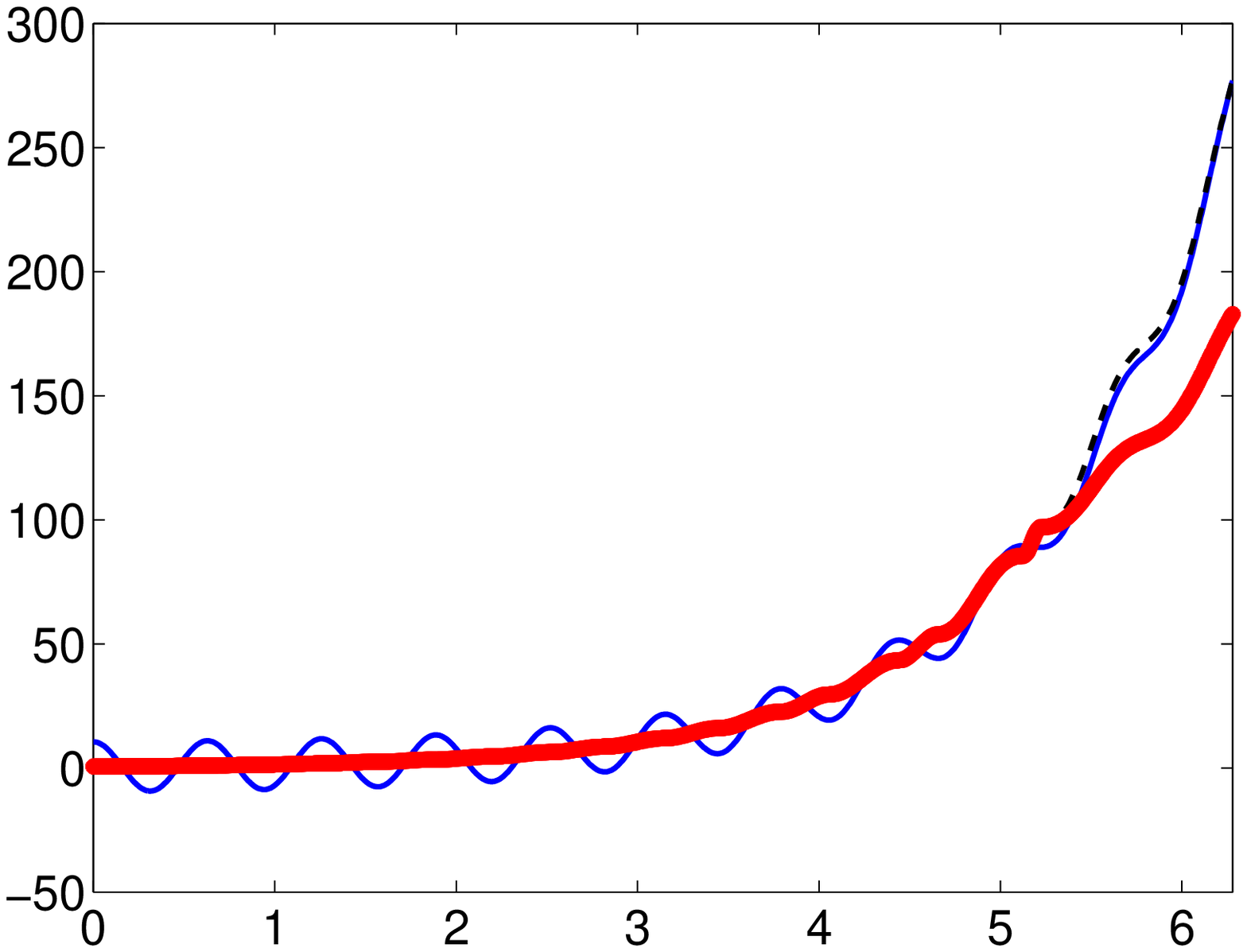}
\caption{(a) The signal is  $f(t_i) =  0.5 e^{t_i} + 50 \cos(10 t_i)$;
the tendency is shown as thick line.  The ITD 
was computed with free boundary conditions at both ends;  
(b) The signal is $f(t_i) =  0.5 e^{t_i}  +10 \cos(10 t_i) $; 
the tendency is shown as a  thick line.  It
was computed with free boundary conditions at both ends;  for comparison,
the first ITD baseline computed 
using a free knot condition on the left, and a clamped knot condition on the right, is shown 
as a dashed line. This mixed boundary condition is designed to capture the end behavior.}
\label{fg.endeffect}
\end{figure}
Supposing the knot conditions are free at both ends, 
 Figure \ref{fg.endeffect}a shows the time series (thin), for $a=50$, as well as one of the baselines that will become the tendency (thick), described in Section~\ref{tendency} below. Now, we decrease $a=10$. We note that 
  the right-most local extremum has moved away significantly from the end of the time interval. The tendency, with both end  knots   free, is the thick line in
 Figure \ref{fg.endeffect}b. The dashed line, on the other hand, was the result of a decomposition with the left end knot free, but the right one clamped: $B^{j+1}_{K^j}=B^j(N)$. The dashed-line tendency in this case is arguably more reasonable.

There are other end effects. We emphasize just one.
 As stated in the Introduction, the ITD decomposition of a signal of specified length will not be necessarily
the same as the same signal with added points to the right, 
say. (This is also the case in the  Empirical Mode Decomposition of \cite{huangemdfirst}). Hence, the outcomes of these methods are by no means
unambiguous when used for extrapolating or forecasting over a time range 
exceeding that of the time series.

\section{The Tendency}
\label{tendency}

Most notions of trend for a time series come equipped with a methodology that  in itself defines the sense in which 
it captures a characteristic of the original signal. 
In order to distinguish our trend from other versions,
we refer to  our analysis as the process of computing the {\it tendency} $\{T(i)\}_{i=1}^N$ of the time series $\{Y(i)\}_{i=1}^N$. The process of determining a tendency for a time series amounts to applying a set of criteria that we define a tendency to have to a collection of time series that are related to the original one. The tendency is thus independent of the manner used to obtain the collection of time series. We use the ITD decomposition process to generate this collection of time series. We use the ITD because it is adaptive, fast, robust, and because the application of a mulitscale diffusion process as a filter captures the spirit of the modeling enterprise, wherein one wants to find characteristics of the signal that are prominent and obtain a complement that could be conceivably well captured by a simple stochastic parametrization.

The tendency, in some informal way, should capture some essential elements of a time series: its inherent time scale structure and the most significant part of its histogram;
the tendency of a strictly monotonic series is the series itself; and
the tendency of a series of constant values is the series itself.

The multiscale structure of a signal can be ascertained qualitatively from the distribution of the locations of its local extrema, and the importance of these local extrema to the total density, estimated by the histogram. This is extracted from projections onto the time axis.  
The distribution of the data 
is encoded in a histogram of the projection of the signal on the vertical axis.  Our two principal
diagnostics extract certain quantitative information from these two aspects of the decomposition.

 \subsection{Horizontal Projection: the Correlation $c^j$}

The measure that most critically determines the choice of the baseline to be the tendency is the 
empirically determined correlation. We have found that the quantity
\[
c^j:=\frac{1-\tilde c^j}{1-\tilde c^1}, \qquad \mbox{where} \, \, \tilde c^j = \mbox{corr}(Y,Y-B^j), \quad j=1,2,..,d-1.
\]
is convenient for graphical depiction of numerical results.

We think of the process of ITD iteration as extraction of the noise-like rotation components to reveal the
part of the signal that carries inherent information found in the signal.  When all this noise has been removed, the next
baseline is declared to be the tendency.  As can be inferred from the analysis  in Section \ref{diffusion}, the correlation between the 
signal and the first rotations should be low, and the correlation between the first baselines and the signal, high. 
Further iterations break up the meaningful component
artificially, and those baselines and rotations should be more correlated.  
Indeed
one can see in Figure~\ref{fg.fbmbaselines}, which is typical in this respect, that 
both $B^j$ and $R^j$ become flatter, and are therefore correlated for a trivial reason.
In our experience, there
is a noticeable downward jump of the  parameter~$c^j$
at a certain~$j$, and after that, not much of a pattern.
The tendency is the a baseline that is still highly correlated with the signal, but one that is not too highly correlated with the rotations. The choice of a suitable baseline for the tendency is 
made less ambiguous by the symmetry statistic  $s^j$ described next.
 
\subsection{Vertical Projection: the Symmetry Statistic $s^j$}
The other determining diagnostic is a measure comparing the
symmetry of the baselines to the symmetry of the
signal.
Define the {\it fluctuation} time series of the signal $Y$ with respect to a signal $T$ (which will be
the candidate tendency) by
$
F(i):= Y(i)-T(i), \quad i=1, 2,...,N.
$
After the ITD decomposition is done,
 the histogram of the fluctuation is computed for each baseline. Each $B^j$ is considered to be
 a potential  tendency. 
 We employ an empirical measure of symmetry,
 namely, the $x$-percentiles, $Pr^j_{x}:=Pr_{x}(F^j)$, from which we define
the symmetry estimator $s^j$ for level $j$ to be
\[
s^j= \frac{Pr^j_{75}-2 \, Pr^j_{50}-Pr^j_{25}}{(Pr^j_{75}-Pr^j_{25})}
\]
  We pick baseline candidates whose associated fluctuation time series
is the most symmetric, $s^j\sim 0$.
In most instances we just compare the 
absolute values
of $s^j$; however, we retain 
sign information as it is sometimes useful in  further
choices between potential tendencies.

\subsection{Supplementary diagnostics}
Symmetry and correlation are the most important properties of our tendency, but we look for confirmation
to two other diagnostics (they are included in the examples below).

\noindent 
{\bf The spread $v^j$:}
For a given $j$, the spread $v^j$ is the unsigned difference between the standard deviation of the baselines and the rotations. These are normalized to the standard deviation of the signal $Y$.
The standard deviation of the baseline is always decreasing.
The spread will reflect certain qualities of the signal and its decomposition. A signal that is random and stationary will have a $v^j$
that remains small throughout the $j$ range. This is especially so for a signal that has all scales, in the sense of having a dense and wide spectrum. If the signal is mutiscale, meaning that its spectrum contains several dense spectral ranges separated by gaps, the 
spread is large and it decreases as~$j$ increases. 
It is typical that the spread reaches zero before~$j$ reaches the last baseline index~$D$. 

\noindent {\bf The Hellinger Distance:}
For two probability densities $p_1$ and $p_2$ on $ \mathbb{R} $, define
$
H(p_1,p_2) = \frac{1}{2}\int [ \sqrt{p_1} - \sqrt{p_2}]^2 = 1- \int \sqrt{p_1p_2}.
$
This is a special case of the {\it Hellinger distance}, 
a measure of the difference between
probability measures.  Since the tendency should capture the most important, non-random, features
of the time series, its Hellinger distance to the pdf of the original signal should be small.
\subsection{Choosing $B^{j^*}=T$, the baseline that becomes the tendency}

In the examples presented in the next section, we follow these steps.

1) By normalization, the correlation parameter is initially equal to~$1$.
Typically, it decreases gradually as~$j$ increases, indicating that the
correlations between baselines and rotations are small.  Often, there 
is a~$j^*$ beyond which~$c^j$ drops significantly, meaning that the rotation and
baseline become correlated.  See Figure~\ref{fg.fbmbaselines}, first panel, below.
The baseline $B^{j^*}$ becomes the candidate tendency.

2) As in Figure~\ref{fg.ocean2}, 
the correlation parameter may sometimes decrease gradually.
In those cases, we may need to choose two or three candidates $B^j$ for
the tendency, and use  the symmetry criterion, i.e. that $s_j$ be close to zero,
to single out~$j^*$.

3) The choice of a particular baseline as tendency can be further tested by
examination of spread and Hellinger distance. Those quantities are shown
in the examples below.

The recent papers~\cite{mbf11,mbf13} have proposed a definition of trend in
terms of the intrinsic mode functions (IMFs) of the EMD.  The number
of zero crossings of IMFs decreases, on average, by a factor of 1/2 per step.
At some point, this pattern breaks down.  The trend is now defined to be
the sum of all the subsequent IMFs.  This criterion is probably related to
our requirement of an increase in correlation between rotation and baseline.
In both approaches, the oscillatory components extracted by the respective
algorithms become spurious modes; they no longer represent noisy fluctuations.

\section{Examples}
\label{examples}

\subsection{Synthetic signals.}
\subsubsection{A stochastic process}
We  analyzed a signal consisting of 512 points from a fractional Brownian motion process with 
Hurst exponent of 0.7.  We pretended  that it consists of a non-random ``carrier"  perturbed by noise.
This is of course not true; the signal is random, but nonetheless, our criteria combine to
pick out the best prospect for a tendency.  They are satisfied only
approximately, but that is the typical situation.

\begin{figure}[hbt]
\centering
(a)\includegraphics[width=6.1cm,height=1.5in]{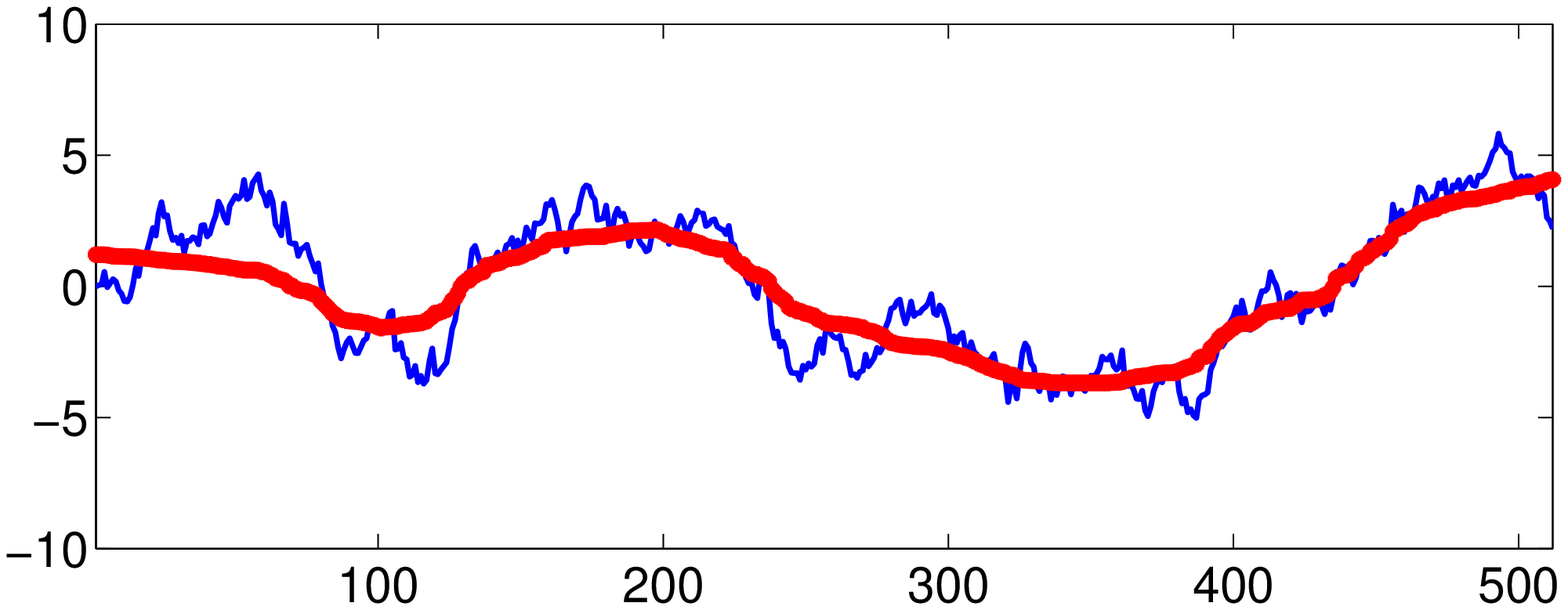}
(b)\includegraphics[width=6.1cm,height=1.5in]{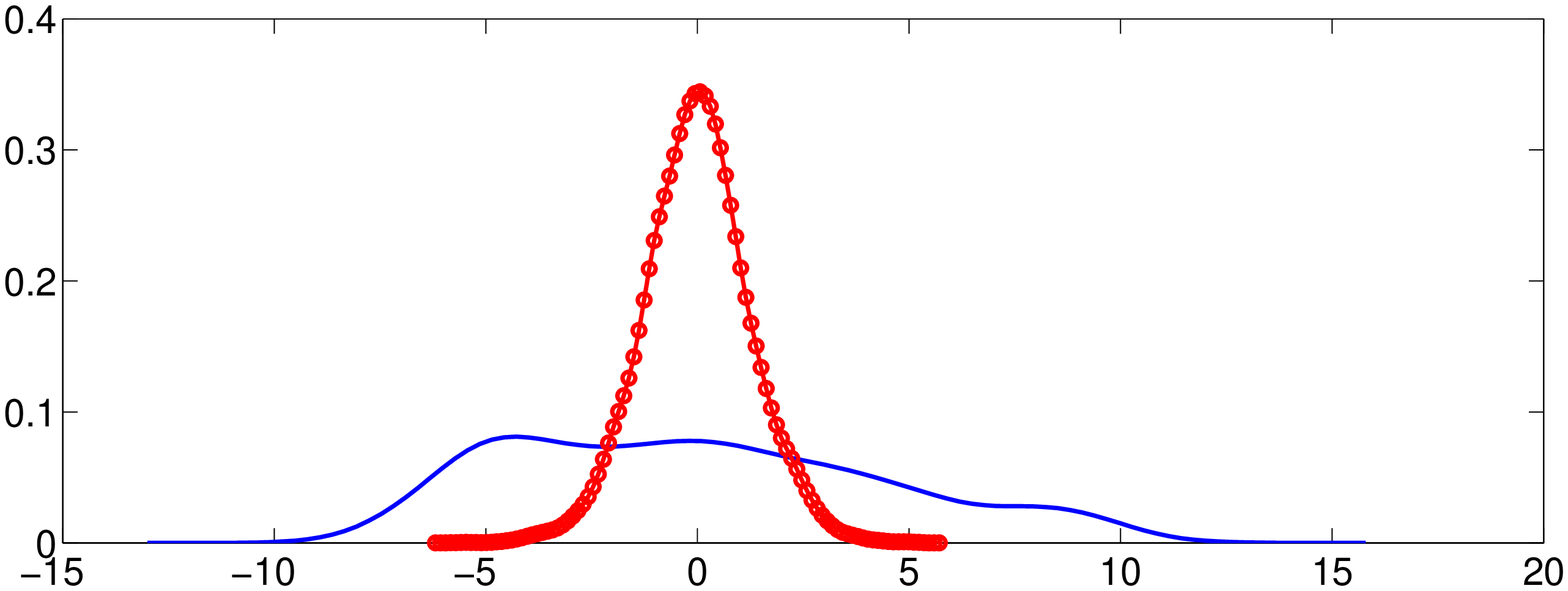}
\caption{The tendency is baseline $B^5$ for the 512-point  $H=0.7$ fractional Brownian motion signal. 
(a) The signal (thin) and the tendency (thick), 
(b) The empirical pdfs of the original signal $Y$ (thin) and
of the fluctuation time series $Y-B^5$ (heavy).
See the text for explanation.}
\label{fg.fbm1}
\end{figure}

\begin{figure}[hbt]
\centering
\includegraphics[width=5in,height=3in]{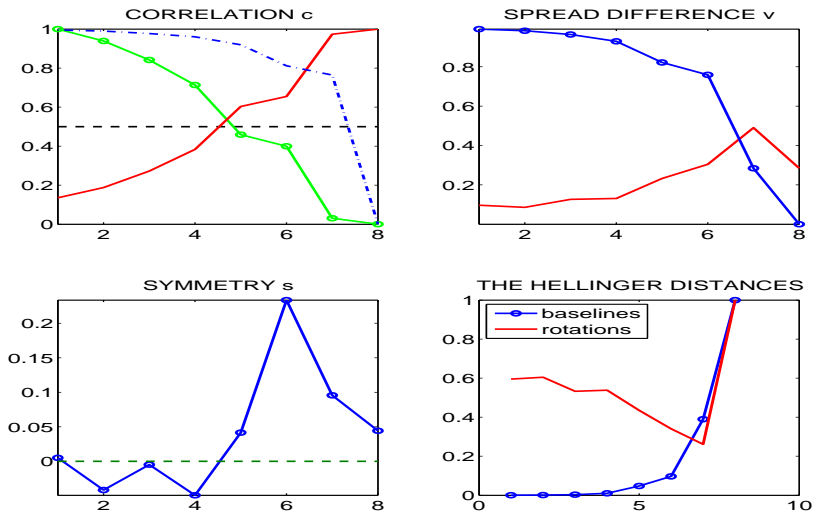}
\caption{Diagnostics for the 512-point  $H=0.7$ fractional Brownian motion signal.
In (a) we plot the $c^j$ as connected dots. The signal has a multiscale nature, as evidenced
by the lighter solid curve, which corresponds to corr$(Y,Y-B^j)$ and the dashed curve which is a plot of corr$(Y,B^j)$.
 In (b)-(d) the baseline data are dotted and rotation data are solid.
See the text for discussion.}
\label{fg.fbm2}
\end{figure}

The tendency appears as the heavy line
in Figure \ref{fg.fbm1}(a). It was determined to be  
the baseline $B^5$, based on the following observations
(see Figure~\ref{fg.fbm2} and Figure~\ref{fg.fbm1}(b)).  The correlation parameter
$c^j$ jumps down at $j=5$.  The symmetry measure
$s^5$ is not close to zero, but
Figure~\ref{fg.fbm1}(b) shows that the fluctuation time series,
$Y-B^5$, nonetheless
has a quite symmetric empirical pdf, with variance smaller than that of the original signal.
Because the correlation jump at~$c^5$ is so pronounced,
we wound up choosing $B^{j^*}=B^5$.

\begin{figure}[hbt]
\centering
\includegraphics[width=2.5in,height=3.5in]{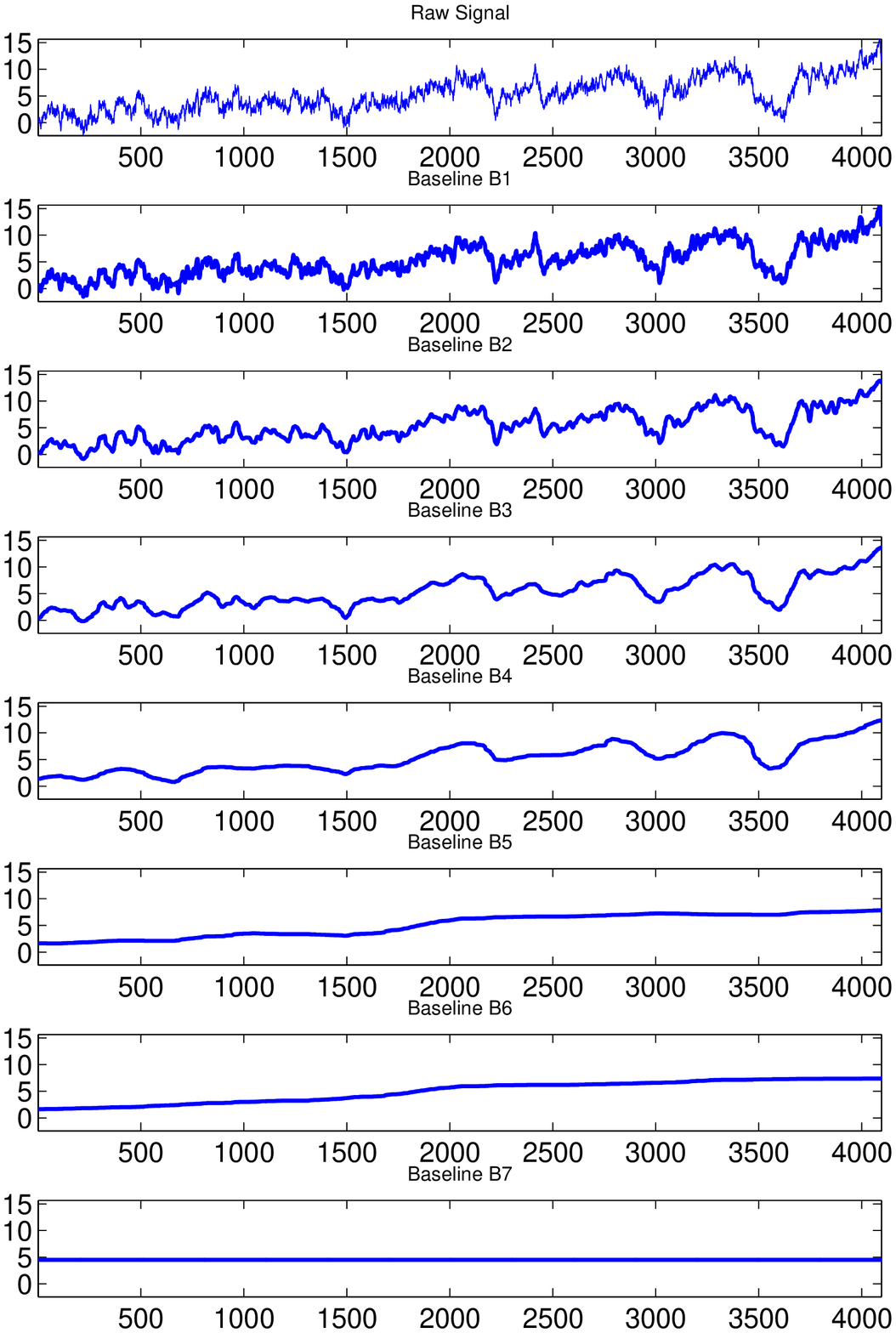}
\includegraphics[width=2.5in,height=3.4in]{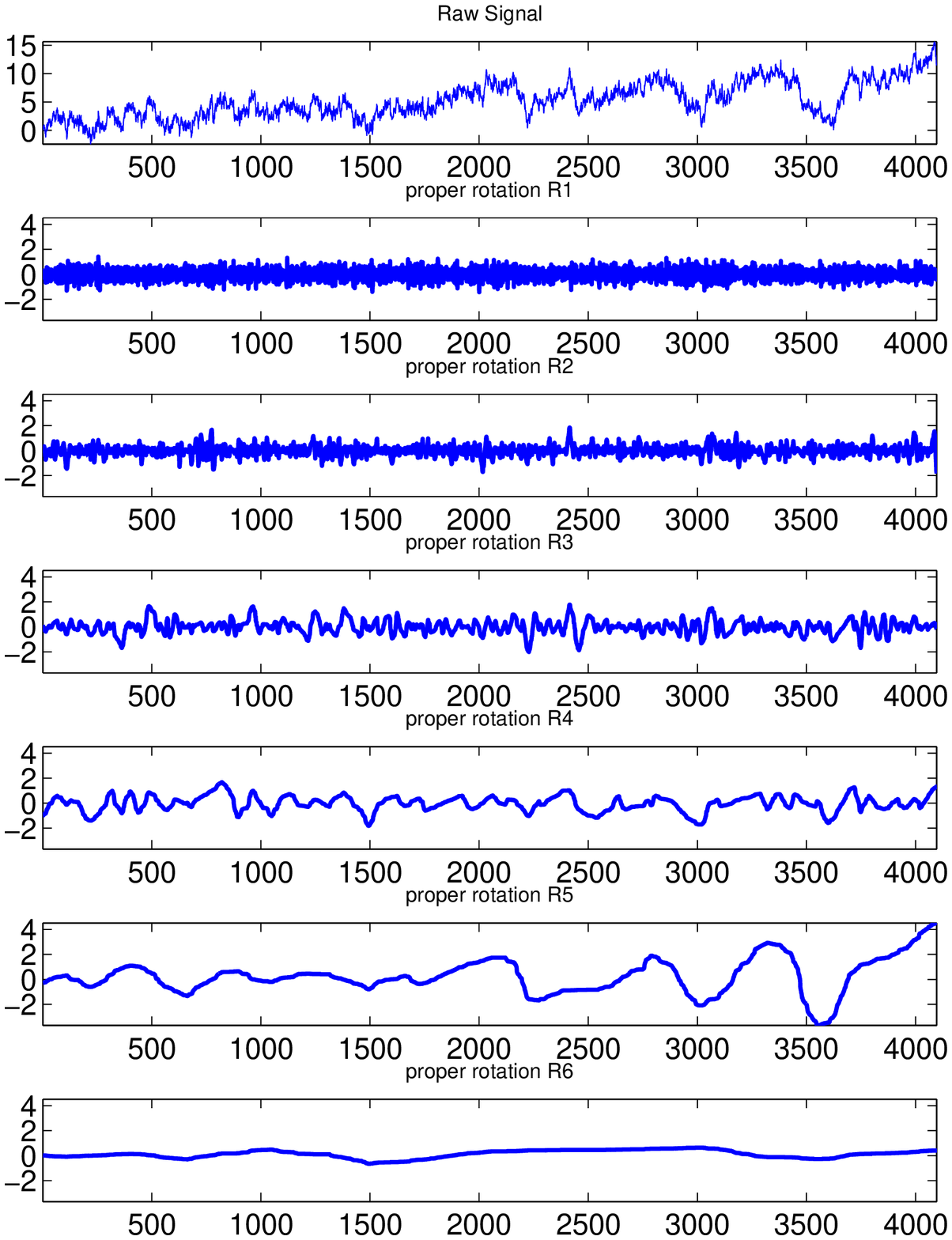}
\caption{The original signal, baselines and rotations
$B^j,R^j$, 
 for the 512-point  $H=0.7$ fractional Brownian motion. The sixth from the top
is the tendency $B^5$.}
\label{fg.fbmbaselines}
\end{figure}

There are eight baselines; it is generally true that
the first baseline is close to the original signal, and the last one
is flat and uninformative.   The  first six
baselines show significant large-scale structures with small-amplitude small scale structures, superimposed.
 This is clear in Figure~\ref{fg.fbmbaselines}.  
 
 The spread $v^j$ (the difference between the top and bottom curves in
 Figure~\ref{fg.fbm2} (b))
 decreases more rapidly from $j=5$ on, indicating that the variances of rotation
 and baseline become more equal; this suggests that they fluctuate on the same
 scale, and the noise has been removed.
 
The Hellinger distance of $B^5$ from the signal is small. The histograms of the
signal and the baseline are similar (they are not pictured here).

It is true that the baseline $B^5$ is
more or less the ``tendency" one would draw by hand.
However, here it is produced algorithmically by the ITD, 
and chosen from the list of candidates by
reasonable quantitative measures.  


\subsubsection{Deterministic Signals}.

\begin{figure}[hbt]
\centering
(a)\includegraphics[width=6.1cm,height=1.1in]{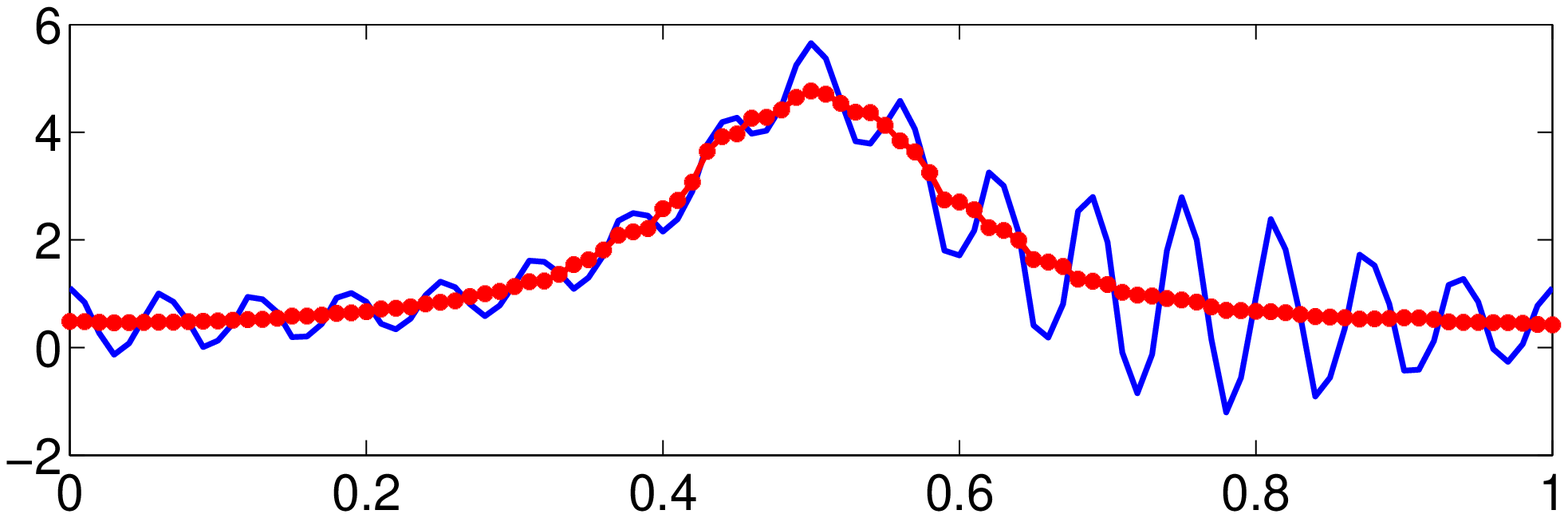}
(b)\includegraphics[width=6.1cm,height=1.1in]{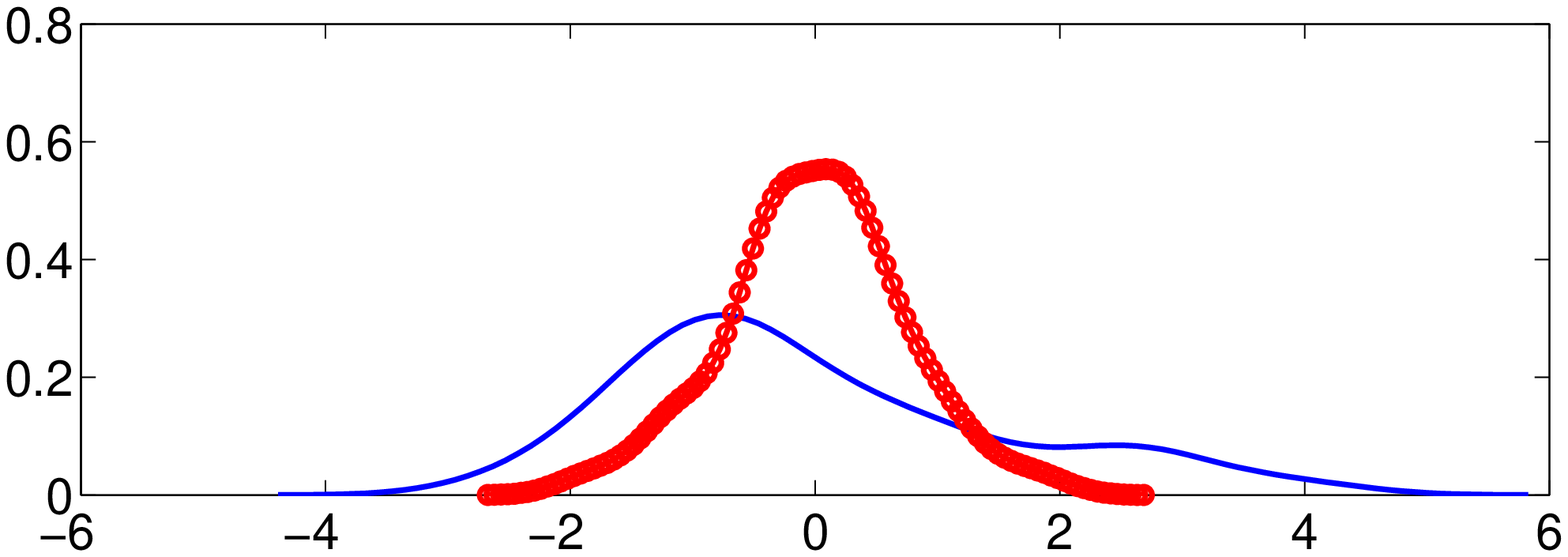}
\caption{Determination of the tendency for the signal given by (\ref{housignal}). (a) The signal (thin) and
the tendency  (heavy);
 (b) The empirical pdfs of the original signal (thin) and of the fluctuating component for the ITD 
 tendency (heavy) }
\label{fg.hou}
\end{figure}

Fully deterministic signals with strong multiscale character are particularly problematic 
for the estimation of trends, when nothing is known about
the underlying structure of the signal.
Here we consider data that have been carefully engineered to have multi-scale
character.
An example of a multiscale signal with challenging qualities is
\begin{equation}
Y(i)=  \frac{1}{1.5 + \sin (2 \pi t_i)} \cos[32 \pi t_i + 0.2 \cos(64 \pi t_i)]+ \frac{1}{(1.2+\cos(2 \pi t_i))}
\label{housignal}
\end{equation}
for $t_i \in$[0:0.0025:1].
This signal was investigated in \cite{houemd}. It was designed to be a test of standard Fourier-based
resolution or wavelet-based multi-resolution techniques. The result of the determination of the 
tendency appears in Figure~\ref{fg.hou}. 
Because the ITD algorithm is based on extraction of
 extrema, even if they are not equally spaced, it is capable of removing the faster oscillations more efficiently 
than a global spectral method, for example. For this example,  the $j=1$ baseline is the chosen tendency. We found that 
the ITD  decomposition was sensitively dependent on the sampling rate. We did not pursue this issue further, other than to 
confirm its existence computationally.

\subsection{Climate Data}

\subsubsection{Ocean temperatures.}
\begin{figure}[hbt]
\centering
(a)\includegraphics[height=1.5in,width=1.5in]{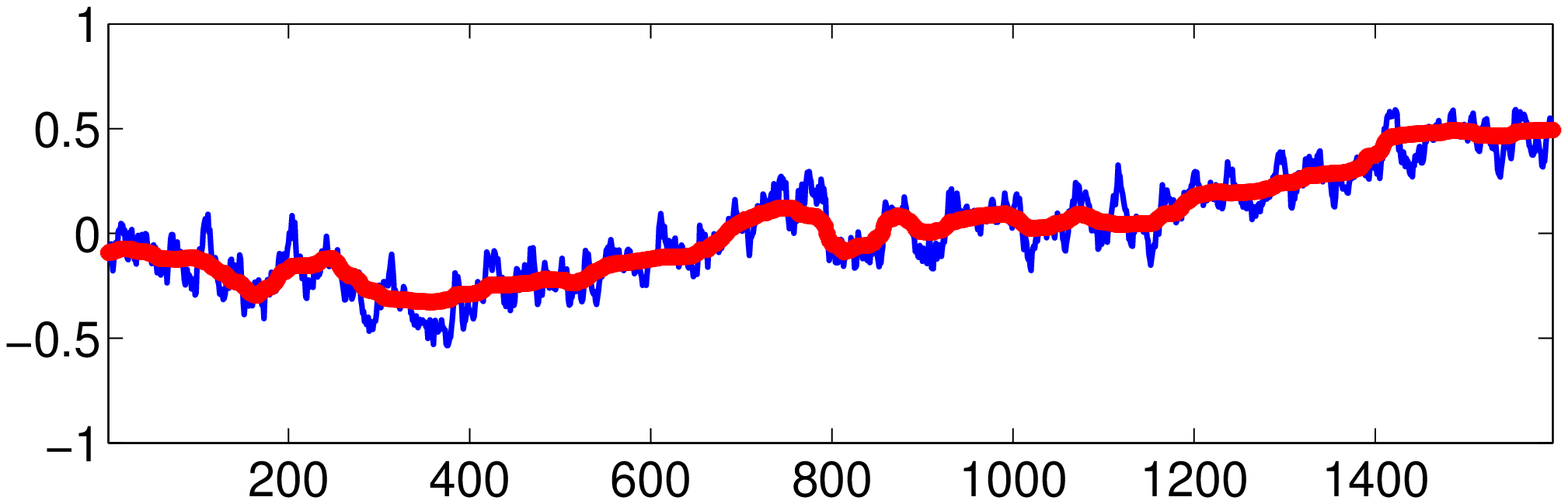}
(b)\includegraphics[height=1.5in,width=1.5in]{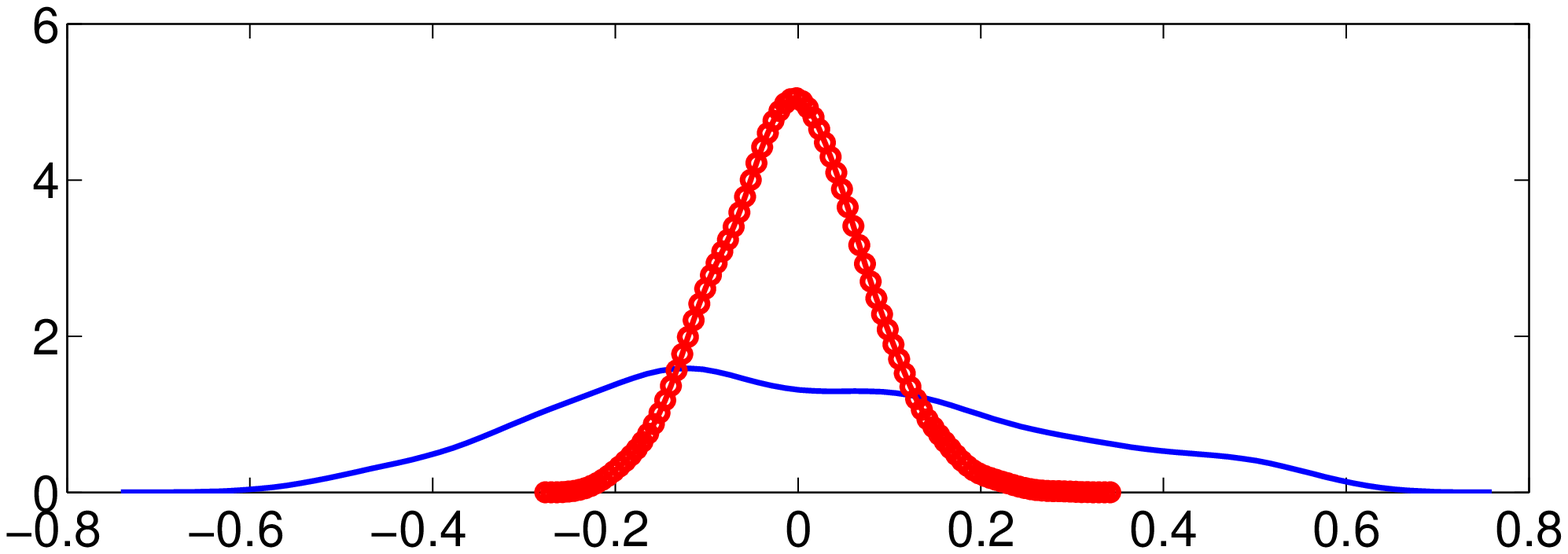}
(c)\includegraphics[height=1.5in,width=1.5in]{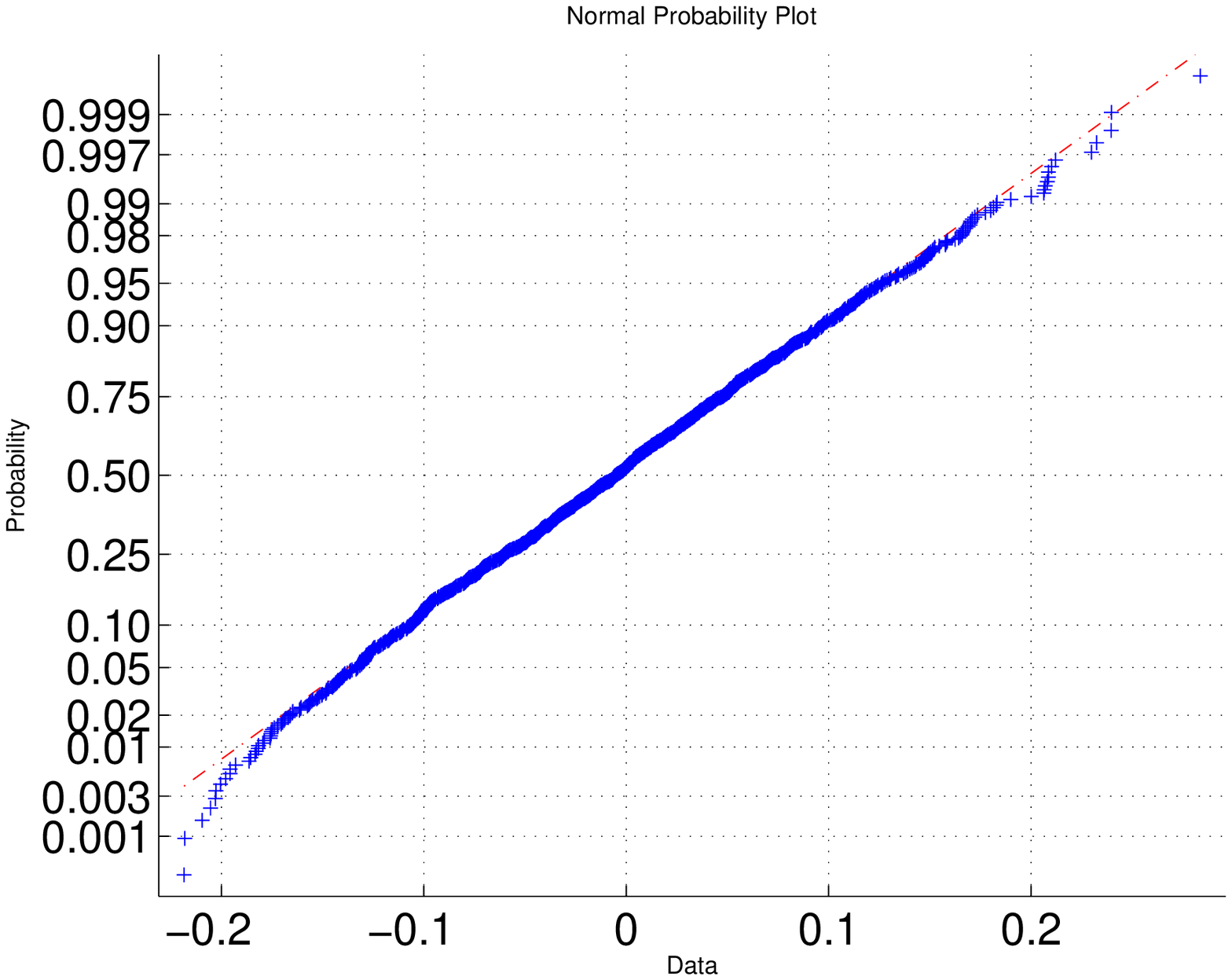}
\caption{(a) Ocean temperature anomalies (thin), in degrees Celsius. The horizontal axis is the number of months, starting with
January of 1880, and ending in December of 2012. The tendency is (dark). (b) Histogram of the signal (thin), histogram of the fluctuation associated with the tendency (dark)  (c) Fit to a normal distribution (dashed) of the fluctuation  associated with the tendency (solid). See also Figure \ref{fg.ocean2}.}
\label{fg.ocean}
\end{figure}

\begin{figure}[hbt]
\centering
\includegraphics[height=2.1in,width=4.2in]{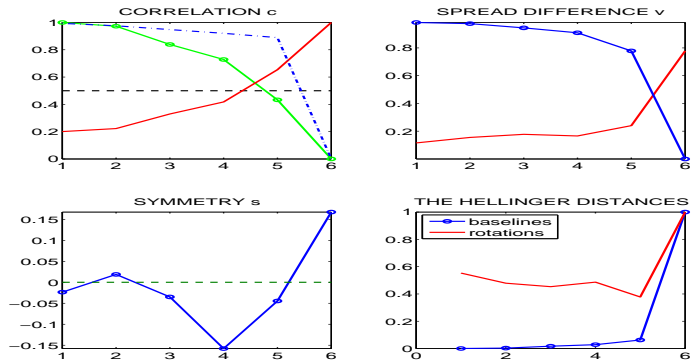}
\caption{Diagnostics for the tendency, accompanying Figure~\ref{fg.ocean}. 
In lexicographic order: $c^j$, $v^j$ baselines (dots), rotations (lines), $s^j$, and the Hellinger distance of the baselines  (dots) and the rotations (lines).}
\label{fg.ocean2}
\end{figure}

We next consider a long time series of monthly ocean 
temperature anomalies dating back to 1880
(available via \\
{\tt ftp://ftp.ncdc.noaa.gov/pub/data/anomalies/monthly.ocean.90S}).
The anomaly signal consists of  fluctuations about the 20th Century average.
The analysis appears in Figures \ref{fg.ocean} and \ref{fg.ocean2}.
In Figure \ref{fg.ocean}a
we display the temperature anomaly (light) and the tendency (dark). The baseline chosen for the tendency corresponds to $j^*=3$. According to the correlation criteria, baselines $j=3$ and $j=4$
are suitable, but the symmetry is higher for $j=3$. See Figure \ref{fg.ocean2}. The diagnostics
indicate that the tendency is suitably close, with regard to the Hellinger distance, to the signal itself.
The spread difference suggests that this is an inherently multiscale signal. The tendency choice
leads to a fluctuation histogram shown in Figure \ref{fg.ocean}b. Its symmetry and fast decay in the tails lead us to compare the fluctuation to a Gaussian. The fit appears in Figure \ref{fg.ocean}c.
\subsubsection{Arizona surface temperature anomalies}
\begin{figure}[hbt]
\centering
(a)\includegraphics[width=6.1cm,height=1.1in]{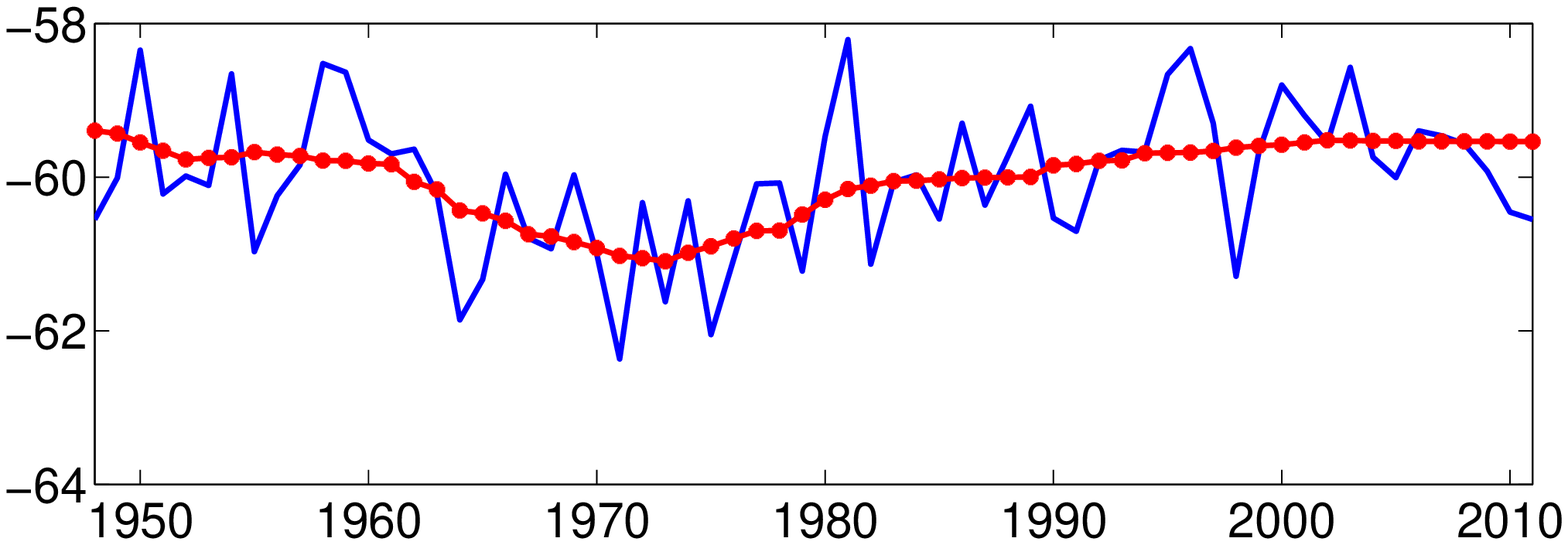}
(b)\includegraphics[width=6.1cm,height=1.1in]{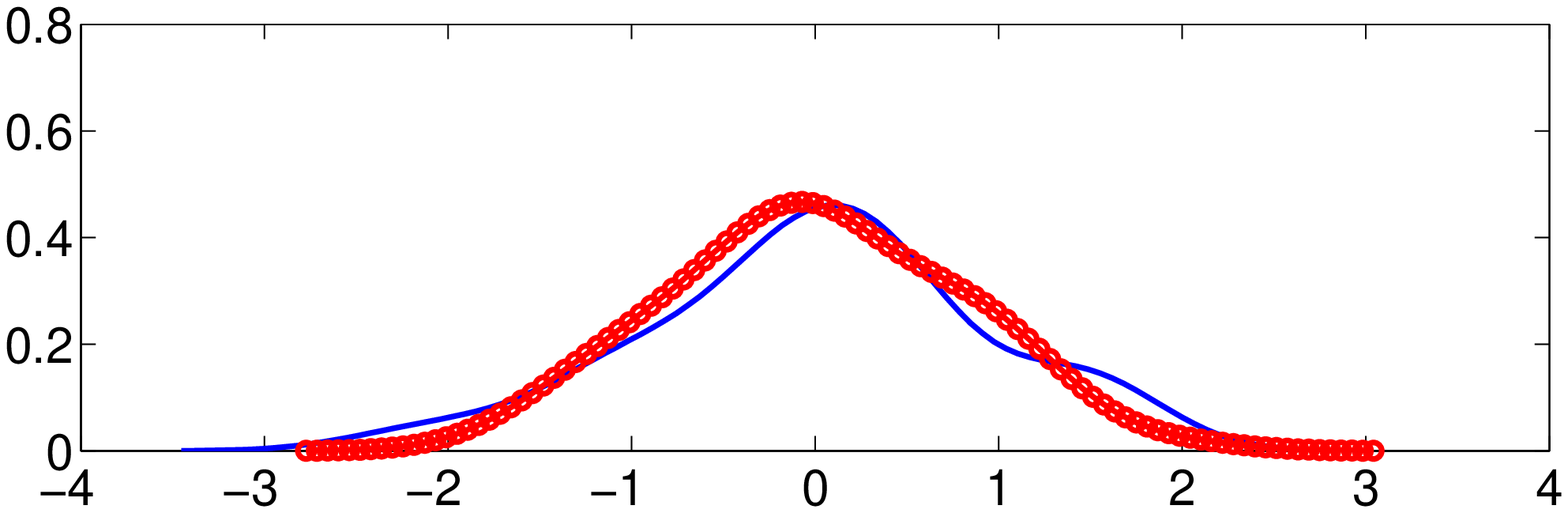}
(c)\includegraphics[width=6.1cm,height=1.1in]{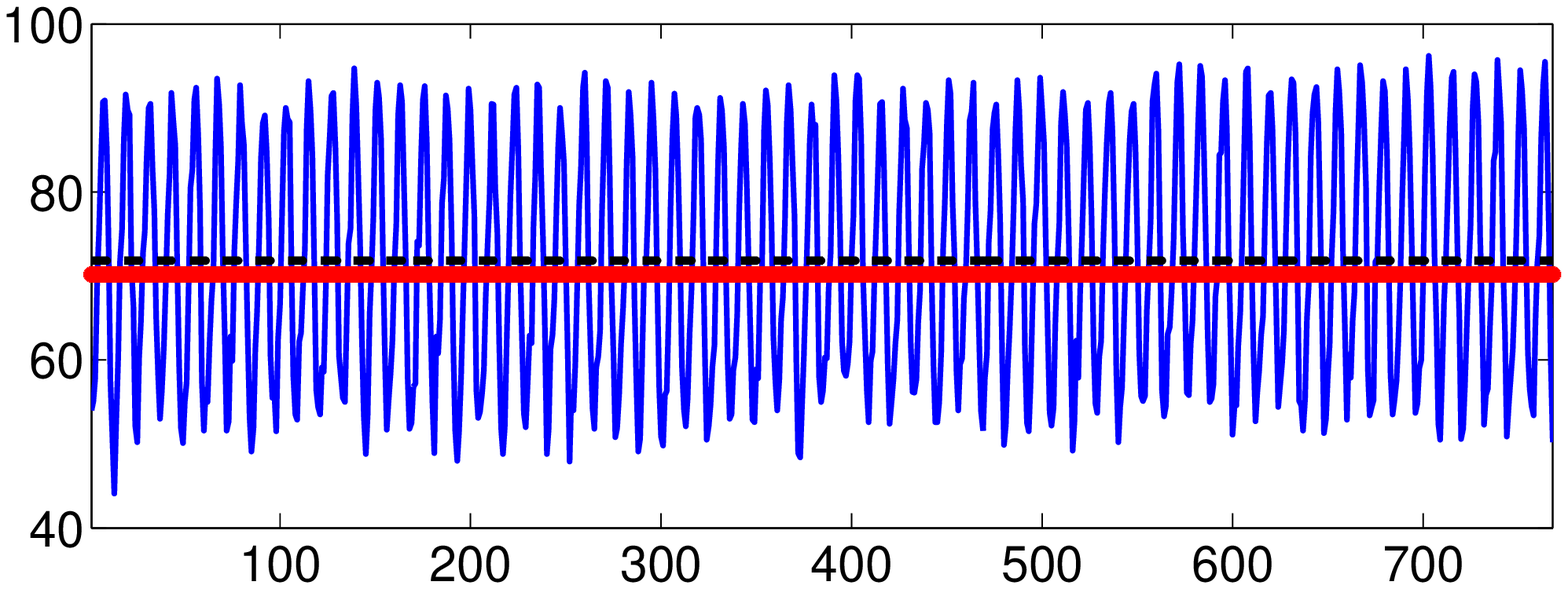}
(d)\includegraphics[width=6.1cm,height=1.1in]{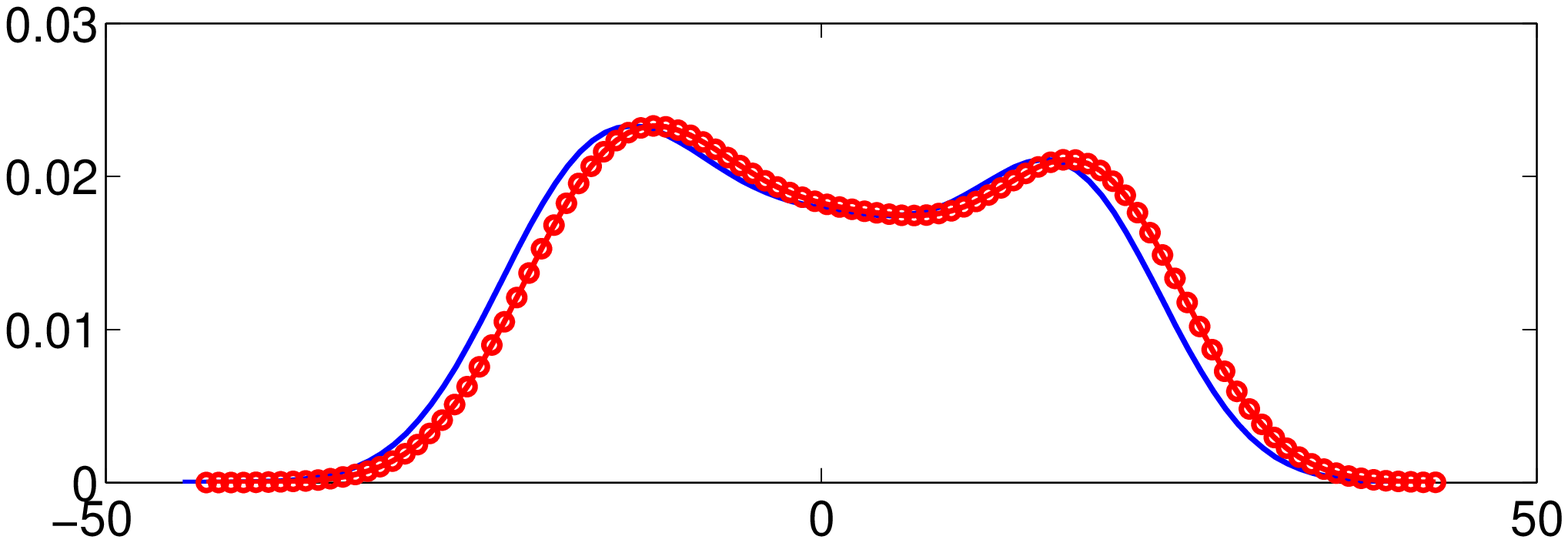}
\caption{Determination of the tendency for the signal given by annual temperature data in Southwest Arizona, from 1948 to 2011. 
The data are drawn from the NOAA/NWS/GISS web site.  
(a) Annually-averaged temperature anomalies. The signal (light), and the tendency (heavy);
 (b) Annually-averaged temperature anomalies. The empirical pdf of the fluctuating component and
 the pdf of the signal (light). (c) The tendency for
 the time series of monthly temperature anomalies. The mean is shown as dotted line.
 (d) The empirical pdfs of the signal and of the fluctuating fields.}
\label{fg.az}
\end{figure}

The annually-averaged temperature in Southwest Arizona for the period 1948-2011 appears in 
Figure~\ref{fg.az}a, with
tendency in red. Figure \ref{fg.az}b shows the corresponding empirical pdfs of the raw data (thin)
and tendency fluctuation (heavy). 
The data can be obtained from  the NOAA National Weather Service
GISS system.  When all 768 monthly records are used for the input time series we obtain the results 
portrayed in  Figure \ref{fg.az}c and d. 
In this case we note that the tendency is consistently below the data mean (dashed),
and as a result the empirical pdf of the tendency fluctuation 
will shift to the right, as compared to the data pdf.  The reason is that the tendency is more sensitive to the asymmetry of the input signal: the lower portion of the signal is far less uniform than the highs. 

\subsubsection{Moscow temperatures.} 
We now apply our method to the time series of July temperatures in Moscow
from 1881 to 2011. The data are found at the NOAA web site
and on the homepage of S.\ Rahmstorf. 
\cite{rahmstorfextremetemps} define a nonlinear trend of this series, and
conclude 
  that the unusually high Moscow summer temperatures of 2010 were
 a result of a gradual increase in the global temperature, rather than 
 being an exceptionally large but otherwise normal fluctuation of the weather.  
Since the data set was relatively
short (131 points), their conclusion was also based on
 expert knowledge of Earth's climate, {\it e.g.}, 
 of climate scales, on which to base the windowing of a 
 moving average calculation, and an assumption of an underlying Gaussian 
 distribution, about a mean, for the temperature data.  
 
 Our goal here is not to focus on the authors' conclusions or 
 methodology. Rather, we are interested in estimating the moving
 average and testing the Gaussianity using only the intrinsic structure of the time series. 

\begin{figure}[hbt]
\centering
(a)\includegraphics[width=2.25in,height=1.5in]{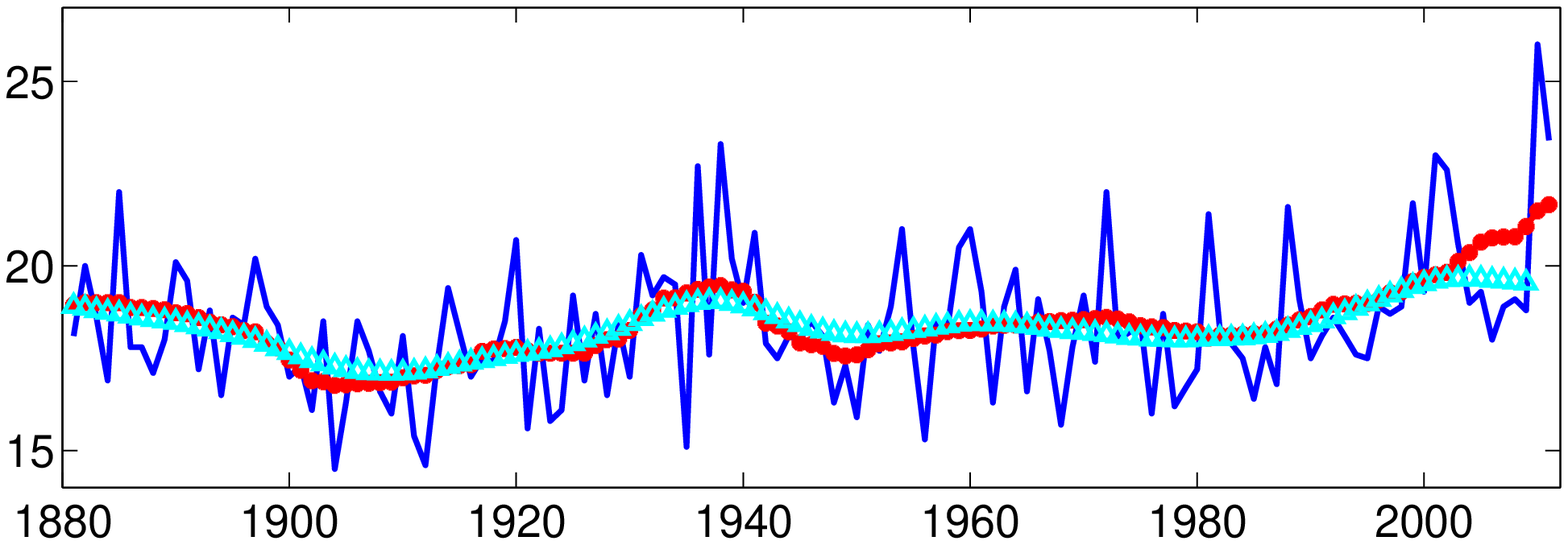}
(b)\includegraphics[width=2.25in,height=1.4in]{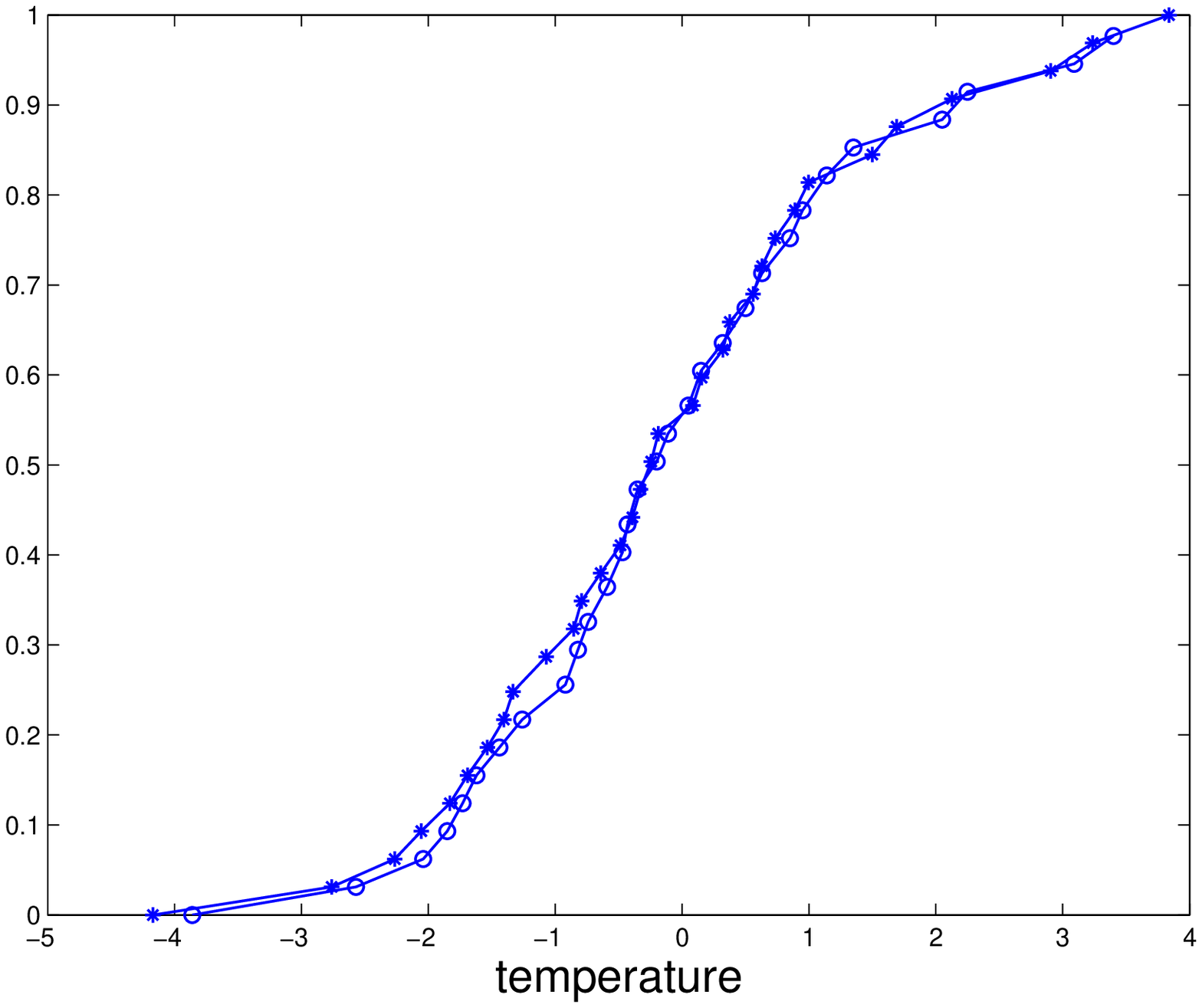}
\caption{July temperatures at the Moscow Station (data from NOAA/NWS/GISS), for 1881-2011. (a) The superposition of the 
real data (light) and the tendency (dark). The Rahmstorff and Coumou 
nonlinear trend (dashed). (b) Empirical cdf of the fluctuations associated with the  tendency (stars) 
the filtered curve  calculated by Rahmstorff and Coumou (circles).}
\label{fg.moscow} 
\end{figure}

In Figure \ref{fg.moscow}a we show
the filtered, or moving, average (which they call ``nonlinear trend line")
 calculated by Rahmstorf and Coumou (light), and the tendency (heavy).
 We followed the procedure described in their paper.
The empirical cumulative distribution functions associated with these data are shown in 
Figure \ref{fg.moscow}b. The tendency was calculated using only the 131 data points, without availing ourselves of knowledge about 
the underlying climate dynamics or 
statistics of the temperature distribution. 
 
\section{Discussion and Conclusions}
\label{disc}
With the aim of addressing the challenge of computing trends for multi-scale signals which 
are not amenable to law-of-large-number arguments we propose a notion of a signal trend which we call the {\it tendency} of the time series. This tendency has been designed to agree closely with an intuitive, rather than a statistical, notion of what a trend for a discrete time series could be.  It is a time series,
of lesser complexity than the original  series, that conveys the most salient features of the histogram and the local time development of the series being analyzed. It emphasizes the  importance
 of more frequent time series values and  more uncommon extremal value locations.

The ITD process yields a decomposition that respects the
 inherent multiscale nature of complex signals. In this regard, 
 the ITD and the EMD yield similar decompositions.  The tendency is found by then
 applying a set of criteria that will identify one of the members of the ITD decomposition as a candidate for the tendency.  Recently, \cite{mbf11} and \cite{mbf13} proposed criteria to determine a trend for a signal using an EMD decomposition. (Another
alternative definition of trend, based upon the EMD decomposition,  is found in  \cite{wuhuanglongpeng}). When the method 
 in \cite{mbf11,mbf13} is applied to the Moscow temperature series, the
 trend is very similar to ours. The criteria proposed for the trend in connection with the EMD analysis consists of examining the ratio of the  energy in the IMF's as well as the ratio of the number of zero crossings.  For random signals it has been observed that the energy  ratios of consecutive IMF's as well as the ratio of the number zero crossings  of consecutive IMF's are very similar. On the other hand, a signal consisting  spectrally-uniform random noise over a structured signal with long timescale features  will yield a decomposition whose ratios will differ at the IMF level corresponding to when the decomposition method no longer picks out mostly noise.  The EMD trend consists of  the sum of the remaining  IMF's.  The tendency will qualitatively agree with the EMD trend for signals of this sort.
 The tendency and the EMD trend will differ when the underlying process that best describes the time series is a random,  intermittent jump process (see \cite{branicki}), and for signals that have underlying trends with significant jumps. 

Until now, very few
analytical properties of the products of the ITD process were known, and those
were derived in the original paper~\cite{itd}. It was established that
 the decomposition method iteratively produces baselines that are guaranteed to have monotonicity when the  signal or the adjacent lower baseline  has local monotonicity. From this we can infer that  the tendency responds to this
by producing a notion of trend that has a high $H^1$-like norm (a norm that combines 
the $\ell^2$-norm of the signal and that of its time-scaled difference values). 
Moreover, if a signal is globally monotonic the tendency would also be globally monotonic. We feel that this is a very strong characteristic of a raw signal that should be included in some notion of a trend for this signal.  In this paper we made some headway in understanding  the ITD process. We studied the decomposition of 
 random stationary signals, numerically and analytically.  
  The numerical results suggested existence of a certain
 scaling universality, and we propose a probabilistic model of the ITD algorithm 
 that exhibits scaling of precisely the type observed experimentally.
 The scaling coefficients obtained by our method are reasonably close to the 
 computed ones, but refinements of the model are needed.
 Because the number of extrema in the baselines scales geometrically, the
number of baselines generated from a signal of length~$N$ is only of order $\log N$,
we suspect that a rigorous explanation  will require an $N\to\infty$ limit, together with some sort
 of renormalization.
We observe that in a formal continuum limit, still for a random signal,
the ITD steps amount to the solution of a diffusion equation; this feature should be
exploited.  It would be interesting to 
extend this analysis to the EMD.

The tendency, we believe, can find use in the analysis of data in which one would like to discern structure in a signal from aspects of the signal that might well be described as random noise of high frequency variability, beyond the standard examples from econometrics. This sort of analysis is commonly done in climate variability,  where one wants to identify aspects of the signal that can be explained by physical models.   Just as other notions of a trend, the tendency requires interpretation. This challenge is presumably one we are willing to accept.

\ack{}
JMR, SV, and DC were supported by NSF grant DMS--1109856. SV also received
support from NSF grant DMS--0807501.
We also acknowledge the support from GoMRI/BP.
The authors wish to thank the anonymous reviewers for suggesting ways to improve the readability of the paper, and further, for alerting us of important literature on finding trends
of time series.
JMR and DC  wish to thank
the Statistics and Applied Mathematics Science Institute (SAMSI) for their
support. Research at SAMSI is 
supported by the NSF. JMR also thanks Prof. Andrew Stuart, Warwick, for stimulating discussions, and  the 
J. T. Oden Fellowship Program at the Institute for Computational Engineering and Sciences (ICES), at The University of  Texas.
\newpage
\appendix
\section{}
\label{append}
\begin{algorithm}
\caption{The ITD Algorithm}
\label{itdalgorithm}
\begin{algorithmic}
\For {$i=1 \to N$}
\State $B^0(i) \gets Y(i)$,
 \EndFor \\
\noindent Find $(\tau^{0},B^{0}_k)$ \, {\it if an extremal value is repeated, pick the right-most of the sequence.}\\
\noindent $K^{0} =\mbox{dim}(\tau^0)$ \\
\noindent $j \gets 0$\\
\While {$K^j \ge 2$}
 \State 
$ B^{j+1}_1 = \frac{1}{2} (B^j_2+B^j_1); \quad  B^{j+1}_{K^j} = \frac{1}{2} (B^j_{K^j-1}+B^j_{K^j});$
\, {\it ''free'' knot conditions at both ends.}
\For { $k=2:K^{j}-1$}
\State 
 \begin{equation}
B^{j+1}_{k}= \frac{1}{2} \left[ B^j_{k-1} +   \frac{(\tau^j_{k}-\tau^j_{k-1})}{(\tau^j_{k+1}-\tau^j_{k-1})}(B^j_{k+1}-B^j_{k-1}) \right] + \frac{1}{2}  B^j_{k}.
\label{bkit}
\end{equation}
\EndFor
\For {$k=1:K^j-1$}
\State
\For { $i=1:N \, \cap \, (\tau^j_k,\tau^j_{k+1}]$}
\State 
\begin{eqnarray}
B^{j+1}(i)  &=& B^{j+1}_k + \frac{(B^{j+1}_{k+1}-B^{j+1}_k)}{(B^j_{k+1}-B^j_k)}(B^j(i)-B^j_k), 
\label{biit} \\
R^{j+1}(i) &=& B^{j}(i) - B^{j+1}(i), 
\label{riit}
\end{eqnarray}

\EndFor
\EndFor

$j \gets j+1$

Find $(\tau^{j})$ \, {\it if an extremal value is repeated, pick the right-most one.}

$K^{j} \gets \mbox{dim}(\tau^{j})$

\EndWhile
\end{algorithmic}
\end{algorithm}
\renewcommand{\baselinestretch}{1.0} 
\vspace{-3cm}

\newpage

\bibliography{itd}
\end{document}